\documentclass[a4paper,11pt]{article}

\usepackage[english]{babel}
\usepackage{graphicx,color,setspace,amsmath}
\usepackage{amssymb}
\usepackage{natbib}
\usepackage{setspace}
\usepackage{mathrsfs}
\usepackage{lscape}
\usepackage{rotating}
\usepackage{verbatim}
\usepackage{amsfonts}  
\usepackage{amsbsy}    
\newcommand{\beq}{\begin{equation}}
\newcommand{\eeq}{\end{equation}}

\newcommand{\beqar}{\begin{eqnarray}}
\newcommand{\eeqar}{\end{eqnarray}}
\newcommand{\bit}{\begin{itemize}}
\newcommand{\eit}{\end{itemize}}
\newcommand{\benum}{\begin{enumerate}}
\newcommand{\eenum}{\end{enumerate}}
\newcommand{\barr}{\begin{array}}
\newcommand{\earr}{\end{array}}
\def\ds{\displaystyle}
\newcommand\eq[1]{(\ref{#1})}
\newcommand{\bfm}[1]{\mbox{\boldmath ${#1}$}}        

\def\XXint#1#2#3{{\setbox0=\hbox{$#1{#2#3}{\int}$}
   \vcenter{\hbox{$#2#3$}}\kern-.5\wd0}}

\def\b0{\mbox{\boldmath $0$}}

\def\bc{\mbox{\boldmath $c$}}

\def\bn{\mbox{\boldmath $n$}}

\def\bp{\mbox{\boldmath $p$}}
\def\bq{\mbox{\boldmath $q$}}

\def\bt{\mbox{\boldmath $t$}}
\def\bu{\mbox{\boldmath $u$}}

\def\bD{\mbox{\boldmath $D$}}

\newcommand{\bsigma}{\mbox{\boldmath $\sigma$}}

\newcommand{\btau}{\mbox{\boldmath $\tau$}}

\newcommand{\bmu}{\mbox{\boldmath $\mu$}}

\def\f0{\ensuremath{\mathbb{O}}}

\newcommand{\Gve}{\varepsilon}

\newcommand{\Gvf}{\varphi}

\newcommand{\Gm}{\mu}

\newcommand{\Gr}{\rho}

\newcommand{\Gs}{\sigma}

\newcommand{\GD}{\Delta}

\newcommand{\GS}{\Sigma}

\newcommand{\BGvf}{\bfm\varphi}

\newcommand{\mD}{\ensuremath{\mathcal{D}}}
\newcommand{\mE}{\ensuremath{\mathcal{E}}}

\newcommand{\mG}{\ensuremath{\mathcal{G}}}

\newcommand{\mI}{\ensuremath{\mathcal{I}}}
\newcommand{\mJ}{\ensuremath{\mathcal{J}}}

\newcommand{\mL}{\ensuremath{\mathcal{L}}}

\def\Im{\mathop{\mathrm{Im}}}

\newcommand{\Reals}{\ensuremath{\mathbb{R}}}

\begin{document}

\centerline{\Large \textbf{Remarks on the energy release rate for an antiplane}}

\centerline{\Large  \textbf{moving crack in couple stress elasticity}}

\begin{center}
L. Morini$^{(1)}\footnote{Corresponding author. Tel.: +39 0461 282583, email address: lorenzo.morini@unitn.it.}$, A. Piccolroaz$^{(1)}$ and 
G. Mishuris$^{(2)}$\\
\end{center}

\centerline{$^{(1)}$\emph{Department of Civil, Environmental and Mechanical Engineering, University of Trento,}}

\centerline{\emph{Via Mesiano 77, 38123, Trento, Italy.}}

\centerline{$^{(2)}$\emph{Institute of Mathematical and Physical Sciences, Aberystwyth University,}}

\centerline{\emph{ Ceredigion SY23 3BZ, Wales, U.K.}}

\vspace{2cm}

\begin{abstract}
This paper is concerned with the steady-state propagation of an antiplane semi-infinite crack in couple stress elastic materials. 
A distributed loading applied at the crack faces and moving with the same velocity of the crack tip is considered, and the influence 
of the loading profile variations and microstructural effects on the dynamic energy release rate is investigated. The behaviour of both 
energy release rate and maximum total shear stress when the crack tip speed approaches the critical speed (either that of the shear waves or that 
of the localised surface waves) is studied. The limit case corresponding to vanishing characteristic scale lengths is addressed both numerically 
and analytically by means of a comparison with classical elasticity results.\\

\end{abstract}

\newpage

\section{Introduction}
Influence of the microstructure on the mechanical behaviour of brittle materials such as ceramics, composites, cellular materials, foams, masonry, bones tissues, glassy and
semicrystalline polymers, has been detected in many experimental analyses \citep{ParkLakes1, Lakes3, WasBev1, BevWhe1}. In particular, relevant size effects have been found when the 
representative scale of the deformation field becomes comparable to the length scale of the microstructure \citep{Lakes1,Lakes2}. These size effects influence strongly the macroscopic
fracture toughness of the materials \citep{RiceFrei1,RiceFrei2}, and cannot be predicted by classical elasticity theory. In order to describe accurately these phenomena, 
generalized theories of continuum mechanics involving characteristic lengths, such as micropolar elasticity \citep{Coss1}, indeterminate couple stress elasticity \citep{Koit1} and 
strain gradient theories \citep{MidEsh1, FleckHutch1, DalWil1}, have been developed and used in many experimental and theoretical studies \citep{RadiGei1, Itou1, Itou2}.

Indeterminate couple stress elasticity theory developed by \cite{Koit1} provides two distinct characteristic length scales for bending and torsion. Moreover, it includes the effects of 
the rotational inertia, which can be considered as an additional dynamic length scale. Full-field solution for steady-state propagating semi-infinite Mode III crack under distributed 
loading has been obtained by means of Fourier transform and Wiener-Hopf analytic continuation technique by \cite{MishPicc1}. A general expression for the dynamic energy release rate (ERR) 
corresponding to the same steady-state antiplane problem has been derived in \cite{MorPicc1}, and the stability of the propagation has been analyzed by means of both maximum total shear stress
\citep{Geo1, Radi1} and energy-based Griffith criterion \citep{Willis1}. In order to investigate how the variation of the applied loading can affect both energy
release rate and maximum total shear stress, in this paper the solution derived in \cite{MishPicc1} is extended considering different distributions for the loading acting on the crack faces 
and moving with the same velocity as that of the crack tip. In particular, the behaviour of the energy release rate in the limiting cases when the crack tip speed approaches the shear waves 
speed or alternatively the Rayleigh-type surface waves speed and when the characteristic scale lengths of the material vanish is studied assuming various 
amplitudes for the loading profile.

The paper starts with a short description of the problem of a semi-infinite Mode III crack steadily propagating in couple stress elastic materials in Section \ref{sec2}, followed by 
an overview of results concerning the dispersive propagation of antiplane surface waves. For both antiplane and in-plane problems, indeterminate couple stress 
theory predicts the existence of surface waves analogous to Rayleigh waves observed in plane classical elasticity \citep{OttRist1}. In the paper, these 
are referred to as couple stress surface waves, and it is demonstrated that the critical maximum value for the crack tip speed introduced in \cite{MishPicc1} and \cite{MorPicc1} coincides with 
the minimum velocity for couple stress surface waves propagation in the material. A velocity range for the crack propagation,  denominated for brevity 
sub-Rayleigh regime, is introduced: in cases where subsonic couple stress surface waves propagation is detected, a maximum crack tip velocity smaller than shear 
waves speed in classical elastic materials $c_s$ is defined and explicitly evaluated as a function of the microstructural parameters, while in cases where the surface 
waves propagation can be only supersonic the limit value for the crack tip speed is given by $c_s$. The analytical full-field solution of the problem is then addressed in Section \ref{sec3} 
using Wiener-Hopf technique \citep{Noble1}. The crack is assumed to propagate in the sub-Rayleigh regime under generalized distributed loading conditions of variable amplitude. 
In Section \ref{sec4}, the dynamic energy release rate is evaluated explicitly by means of the method developed by \cite{Freund2} and extended by \cite{Geo1}, \cite{MorPicc1} and 
\cite{GourPicc} to static and dynamic problems in couple stress elasticity. 

The effects of the microstructure as well as the influence of the loading profile gradients on displacements, stress fields, maximum total shear stress and energy release rate are illustrated 
and discussed by means of several numerical examples in Section \ref{sec5}. A strong localization of the applied loading around a maximum near to the crack tip is not associated with 
to higher levels of the shear traction and to a larger crack opening. This behaviour, detected by maximum total shear stress analysis, means that in couple stress elastic materials the action of loading forces concentrated 
near to the crack tip is {\em{shielded}} by the microstructure. This {\em{shielding}} effect is confirmed also by the energy release rate analysis. It is shown indeed that the energy release rate decreases as the applied loading 
is more and more localized near the crack tip.

The behaviour of the energy release rate shows that if the distance between the position of application of the maximum loading and the crack tip grows, in presence of couple stress 
more energy is provided for propagating the crack at constant speed with respect to the classical elastic case, and then the fracture propagation is favored. Also this
{\it weakening} effect is due to the microstructural contributions, and it is in agreement with the results detected in \cite{GourGeo1} for plane strain crack problems 
under concentrated shear loading. Numerical results illustrate also that, when the crack tip speed approaches the shear waves speed in classical elastic materials or alternatively 
the couple stress surface waves speed, the energy release rate assumes a finite limit value depending on the microstructural parameters. Conversely,
if the characteristic lengths vanish, for any arbitrary loading profile the value of the energy release rate becomes identical to that of the classical elastic case. 
This is an important proof of the fact that, if the microstructural effects are negligible, the material behaviour is identical to that of a classical elastic body for what concerns 
crack propagation. This result, observed in all the proposed numerical examples, is validated by means of the analytical evaluation of the limit of the energy release rate for vanishing 
characteristic lengths reported in Section \ref{sec6}. In this Section, indeed, it is demonstrated that, if the characteristic lengths vanish, 
for any arbitrary applied loading the energy release rate for couple stress materials tends to the energy release rate associated to an antiplane steady-state crack in classical elasticity.

\section{Problem formulation}
\label{sec2}

A Cartesian coordinate system $(0, x_1, x_2, x_3)$ centered at the crack-tip at time $t = 0$ is assumed.  The micropolar behaviour of the material is
described by the indeterminate theory of couple stress elasticity \citep{Koit1}. The non-symmetric Cauchy stress tensor $\bt$ can be decomposed into 
a symmetric part $\bsigma$ and a skew-symmetric part $\btau$, namely $\bt = \bsigma + \btau$. The reduced tractions vector $\bp$ and couple stress tractions 
vector $\bq$ are defined as
\beq
\label{reduced}
\bp = \bt^T \bn + \frac{1}{2} \nabla \mu_{nn} \times \bn, \quad \bq = \bmu^T \bn - \mu_{nn} \bn,
\eeq
where $\bmu$ is the couple stress tensor, $\bn$ denotes the outward unit normal and $\mu_{nn} = \bn \cdot \bmu \bn$. For the dynamic antiplane problem, stresses 
and couple stresses can be expressed in terms of the out-of plane displacement $u_3$:
\beq
\label{sigma}
\sigma_{13} = G \frac{\partial u_{3}}{\partial x_1}, \quad \sigma_{23} = G \frac{\partial u_{3}}{\partial x_2},
\eeq
\beq
\label{tau}
\tau_{13} = -\frac{G\ell^2}{2} \Delta \frac{\partial u_{3}}{\partial x_1} + \frac{J}{4} \frac{\partial \ddot{u}_{3}}{\partial x_1}, \quad
\tau_{23} = -\frac{G\ell^2}{2} \Delta \frac{\partial u_{3}}{\partial x_2} + \frac{J}{4} \frac{\partial \ddot{u}_{3}}{\partial x_2},
\eeq
\begin{gather}
\label{mu}
\mu_{11} = -\mu_{22} = G\ell^2(1 + \eta) \frac{\partial^2 u_{3}}{\partial x_1 \partial x_2}, \quad
\mu_{21} = G\ell^2 \left(\frac{\partial^2 u_{3}}{\partial x_2^2} - \eta \frac{\partial u_{3}}{\partial x_1^2}\right), \nonumber \\
\mu_{12} = -G\ell^2 \left(\frac{\partial^2 u_{3}}{\partial x_1^2} - \eta \frac{\partial^2 u_{3}}{\partial x_2^2}\right).
\end{gather}
where $\Delta$ denotes the Laplace operator, $J$ is the rotational inertia, $G$ is the elastic shear modulus, $\ell$ and $\eta$ the couple stress parameters, with $-1 < \eta < 1$. 
Both material parameters $\ell$ and $\eta$ depend on the microstructure and can be connected to the material characteristic lengths in bending and in torsion \citep{Radi1}, 
namely $\ell_b = \ell/\sqrt{2}$ and $\ell_t = \ell \sqrt{1 + \eta}$. Typical values of $\ell_b$ and $\ell_t$ for some classes of materials with microstructure can be found in \cite{Lakes1, Lakes2}.

Substituting expressions \eq{sigma}, \eq{tau} and \eq{mu} in the dynamic equilibrium equations \citep{MishPicc1}, the following equation of motion is derived:
\beq
\label{motion}
G \Delta u_3 - \frac{G\ell^2}{2} \Delta^2 u_3 + \frac{J}{4} \Delta \ddot{u}_3 = \rho \ddot{u}_3.
\eeq

\subsection{Steady-state crack propagation}
\label{subsec11}

We assume that the crack propagates with a constant velocity $V$ straight along the $x_1$-axis and is subjected to reduced force traction $p_3$
applied on the crack faces, moving with the same velocity $V$, whereas reduced couple traction $q_1$ is assumed to be zero \citep{Geo1},
\beq
\label{loadcond}
p_3(x_1,0^\pm,t) = \mp \tau(x_1 - Vt), \quad q_1(x_1,0^\pm,t) = 0, \quad \text{for} \quad x_1 - Vt < 0.
\eeq
We also assume that the function $\tau$ decays at infinity sufficiently fast and it is bounded at the crack tip. These requirements
are the same requirements for tractions as in the classical theory of elasticity.\\ It is convenient to introduce a moving framework $X = x_1 - Vt$, $y = x_2$. By 
assuming that the out of plane displacement field has the form $u_3(x_1,x_2,t) = w(X,y)$, then the equation of motion (\ref{motion}) writes:
\beq
\label{pde}
\left(1 - m^2\right) \frac{\partial^2 w}{\partial X^2} + \frac{\partial^2 w}{\partial y^2} -
\frac{\ell^2}{2}\left(1 - 2m^2h_0^2\right)\frac{\partial^4 w}{\partial X^4} -
\ell^2\left(1 - m^2h_0^2\right)\frac{\partial^4 w}{\partial X^2 \partial y^2} - \frac{\ell^2}{2} \frac{\partial^4 w}{\partial y^4} = 0,
\eeq
where $m = V/c_s$ is the crack velocity normalized to the shear waves speed $c_s$, and $h_0=\sqrt{J/4\rho}/\ell$ is the normalized rotational inertia defined in \cite{MishPicc1}.\\ 
According to (\ref{reduced}), the non-vanishing components of the reduced force traction and reduced couple traction vectors along the crack line $y = 0$, where
$\bn = (0, \pm1, 0)$, can be written as
\beq
\label{reduced2}
p_3 = t_{23} + \frac{1}{2} \frac{\partial \mu_{22}}{\partial X}, \quad q_1 = \mu_{21},
\eeq
respectively. By using (\ref{sigma})$_2$, (\ref{mu})$_{1,2}$, (\ref{tau})$_2$, and (\ref{reduced2}), the loading conditions (\ref{loadcond}) on the upper crack
surface require the following conditions for the function $w$:
\begin{gather}
\frac{\partial w}{\partial y} - \frac{\ell^2}{2}\frac{\partial}{\partial y}\left[(2 + \eta - 2m^2h_0^2)\frac{\partial^2 w}{\partial X^2} +
\frac{\partial^2 w}{\partial y^2}\right] =
-\frac{1}{G} \tau(X), \nonumber \\
\frac{\partial^2 w}{\partial y^2} - \eta \frac{\partial^2 w}{\partial X^2} = 0, \quad \text{for} \quad X < 0, \quad y = 0^+.
\label{bc1}
\end{gather}
Ahead of the crack tip, the skew-symmetry of the Mode III crack problem requires
\beq
\label{bc2}
w = 0, \quad \frac{\partial^2 w}{\partial y^2} - \eta \frac{\partial^2 w}{\partial X^2} = 0, \ \text{for} \ X > 0, \ y = 0^+.
\eeq
Note that the ratio $\eta$ enters the boundary conditions (\ref{bc1})-(\ref{bc2}), but it does not appear into the governing PDE (\ref{pde}).

\subsection{Preliminary analysis on couple stress surface waves propagation}
\label{subsec12}

In couple stress elastic materials the existence of surface waves has been demonstrated for both in-plane and antiplane problems 
\citep{OttRist1}. Considering a material occupying the upper half-plane under antiplane deformations, the solution of the governing equation \eq{motion} 
is assumed in the form:
\beq
\label{RW}
u_3(x_1, x_2, t)=W(x_2)e^{i(kx_1-\omega t)}, \quad x_1\geq0,
\eeq
where $W$ is the amplitude, k is the wave number and $\omega$ the radian frequency. Substituting \eq{RW} into \eq{motion} the following ODE is obtained:
\beq
\label{RW_ODE1}
W^{''''}-\frac{2}{\ell^2}\left[k^2\ell^2+\left(1-\frac{\omega^2}{\theta^2}\right)\right]W^{''}+\frac{2}{\ell^2}\left[\frac{k^4\ell^2}{2}+\left(1-\frac{\omega^2}{\theta^2}\right)k^2
-\frac{\omega^2}{c_s^2}\right]W=0,
\eeq
where $c_s=\sqrt{G/\rho}$ is the shear wave speed for classical elastic materials, $\theta=\sqrt{4G/J}$ and the superscript $'$ indicates the derivative with respect to $x_2$ 
variable. Equation \eq{RW_ODE1} can be rewritten in the form 
\beq
\label{RW_ODE2}
W^{''''}-\frac{2}{\ell^2}\left[1+\left(\frac{1}{m_R^2}-h_0^2\right)\frac{\omega^2\ell^2}{c_s^2}\right]W^{''}+\frac{1}{\ell^4}\left[\left(\frac{1}{m_R^2}-2h_0^2\right)\frac{\omega^4\ell^4}{m_R^2c_s^4}
-2\left(1-\frac{1}{m_R^2}\right)\frac{\omega^2\ell^2}{c_s^2}\right]W=0,
\eeq
where $m_R=v_R/c_s$, $v_R=\omega/k$ is the couple stress surface waves speed and $h_0=c_s/\theta\ell=\sqrt{J/4\rho}/\ell$ is the normalized rotational inertia introduced 
in the previous section. Equation \eq{RW_ODE2} admits the following bounded solution in the upper half-plane, vanishing for $x_2 \rightarrow +\infty$
\beq
\label{RW_W}
W(x_2)=Ae^{-\alpha(\omega, m_R)x_2/\ell}+Be^{-\beta(\omega, m_R)x_2/\ell}, \quad \mbox{for} \ x_2>0,
\eeq
where
\begin{eqnarray}
\alpha(\omega, m_R) & = & \sqrt{1-\left(h_0^2-\frac{1}{m_R^2}\right)\frac{\omega^2\ell^2}{c_s^2}+\chi(\omega)}= \sqrt{1+\left(1-h_0^2m_R^2\right)k^2\ell^2+\chi(k, m_R)},\\
\beta(\omega, m_R)  & = & \sqrt{1-\left(h_0^2-\frac{1}{m_R^2}\right)\frac{\omega^2\ell^2}{c_s^2}-\chi(\omega)}=\sqrt{1+\left(1-h_0^2m_R^2\right)k^2\ell^2-\chi(k, m_R)},\label{RW_eigen}
\end{eqnarray}
\beq
\label{RW_chi}
\chi(\omega)=\sqrt{1+2(1-h_0^2)\frac{\omega^2\ell^2}{c_s^2}+h_0^4\frac{\omega^4\ell^4}{c_s^4}}=\sqrt{1+2(1-h_0^2)m_R^2k^2\ell^2+h_0^4m_R^4k^4\ell^4}.
\eeq
Similarly to the procedure commonly carried out for studying Rayleigh waves in classical elasticity, traction-free boundary conditions are imposed at the free surface:
\beq
\label{RW_bound1}
p_2(x_1,0^+,t)=0, \quad q_1(x_1,0^+,t)=0, \quad \mbox{for} \ -\infty<x_1<\infty,
\eeq
by using relations \eq{sigma}, \eq{tau}, \eq{mu} together with expression \eq{RW}, equation \eq{RW_bound1} becomes
\beq
\label{RW_bound2a}
W^{'}(0)-\frac{\ell^2}{2}\left[-\frac{\omega^2}{c_s^2m_R^2}(2+\eta-2h_0^2m_R^2)W^{'}(0)+W^{'''}(0)\right]=0,
\eeq
\beq
\label{RW_bound2b}
W^{''}(0)+\frac{\eta \omega^2}{c_s^2m_R^2}W(0)=0.
\eeq
Substituting expression \eq{RW_W} into equations \eq{RW_bound2a} and \eq{RW_bound2b}, the following system of two algebraic equations for the unknown constants $A$ and $B$ is derived
\beq
\label{RW_matrix}
\bD(m_R,\omega)\bc=0,
\eeq
where $\bc=(A, B)^T$ and the matrix $\bD$ is given by
$$
\bD(m_R,\omega)=\left[
\begin{array}{cc}
\alpha^3-\alpha\left(2-\cfrac{\omega^2\ell^2}{c_s^2m_R^2}\left(2+\eta-2h_0^2m_R^2\right)\right)  &   \beta^3-\beta\left(2-\cfrac{\omega^2\ell^2}{c_s^2m_R^2}\left(2+\eta-2h_0^2m_R^2\right)\right)         \\
                                                                        &                                                                                 \\
\alpha^2+\eta\cfrac{\omega^2\ell^2}{m_R^2c_s^2}                          &    \beta^2+\eta\cfrac{\omega^2\ell^2}{m_R^2c_s^2} 
\end{array}
\right],
$$
the system \eq{RW_matrix} possesses non-trivial solutions only if
\beq
\label{RW_disp}
\mD(m_R,\omega) = \det \bD(m_R,\omega)=0.
\eeq
Expression \eq{RW_disp} is the dispersion relation for antiplane couple stress surface waves, and the propagation velocity corresponding 
to a given value of the frequency $\omega$ or alternatively of the wave vector $k$ can be evaluated by solving this equation.

\begin{figure}[!htcb]
\centering
\includegraphics[width=160mm]{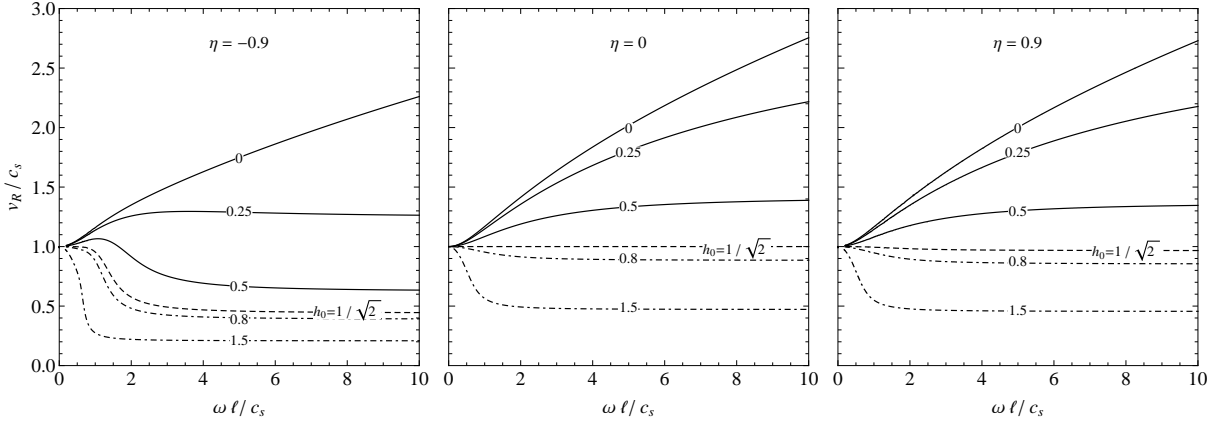}
\caption{\footnotesize Variation of the normalized Rayleigh waves speed with the normalized frequency.}
\label{dispOm}
\end{figure}

\begin{figure}[!htcb]
\centering
\includegraphics[width=160mm]{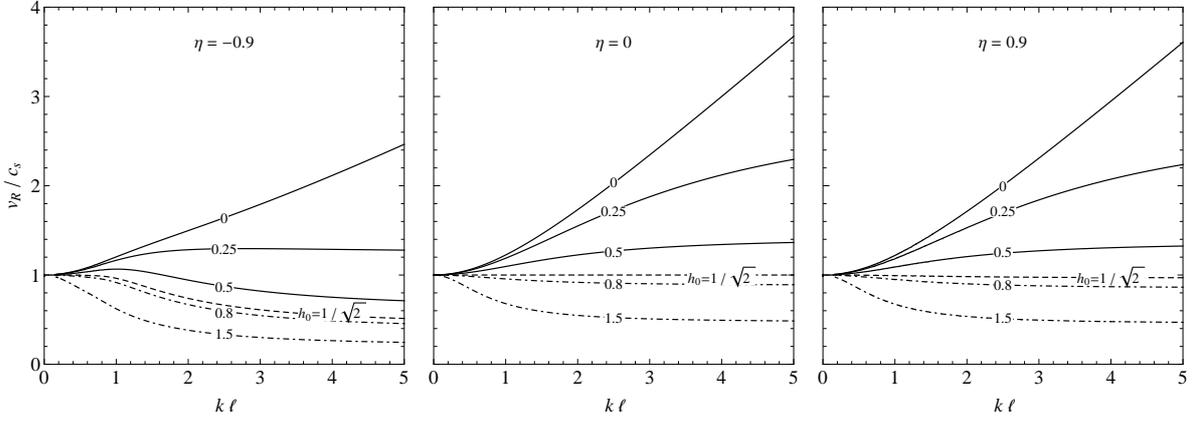}
\caption{\footnotesize Variation of the normalized Rayleigh waves speed with the normalized wave vector.}
\label{dispKl}
\end{figure}
\noindent

The normalized wave speed $m_R=v_R/c_s$ is shown in Figs.\ \ref{dispOm} and \ref{dispKl} as a function of the normalized frequency $\omega\ell /c_s$ and the normalized wave number $k\ell$, 
respectively. Different values for the characteristic parameter $\eta$ and for the normalized rotational inertia $h_0$ have been considered. 

For small values of the rotational inertia, the value of the couple stress surface waves speed is always greater than the shear waves velocity in classical elastic materials, and then the 
couple stress surface waves propagation is supersonic for any value of the wave number and frequency. In particular, for the case of vanishing rotational inertia $h_0=0$, 
the wave propagation is dispersive and supersonic with monotonically increasing speed, as it as been detected in \cite{OttRist1} and \cite{AskAif1}. As the rotational inertia increases, 
the phase speed behaviour changes: the values of $v_R$ may become smaller then $c_s$, and it decreases with the frequency and the wave number until a limit value corresponding to $m_R<1$
and depending by $h_0$ and $\eta$ is reached. This means that for large values of the rotational inertia and high frequencies the couple stress surface waves propagation 
becomes subsonic, and a minimum value for the phase speed is individuated for  $\omega\rightarrow \infty$. 

\begin{figure}[!htcb]
\centering
\includegraphics[width=140mm]{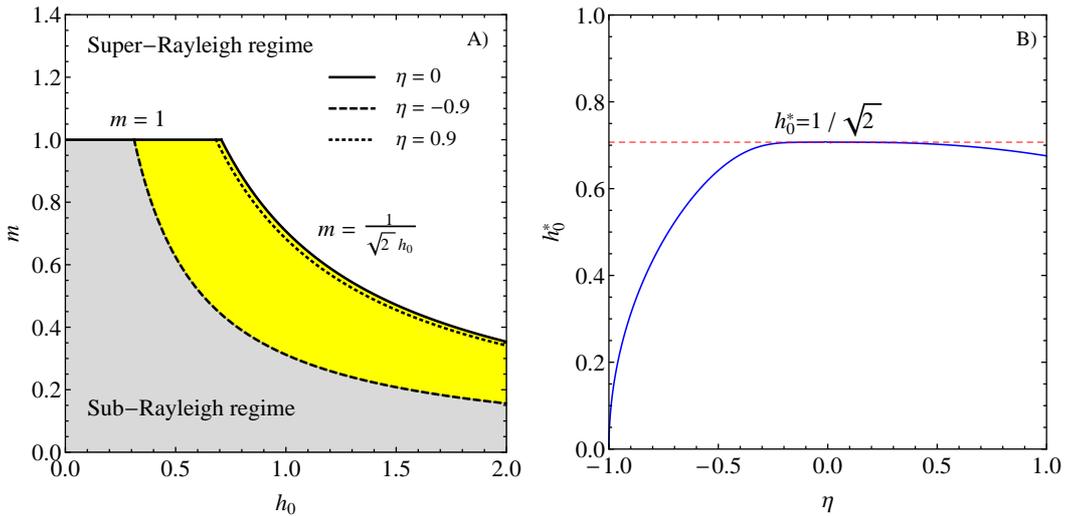}
\caption{\footnotesize A): Sub-Rayleigh and super-Rayleigh regimes in the $m-h_0$ plane. The continuous line coincides with the transition between subsonic and supersonic ranges. B): Variation of 
$h_0^*$ as a function of $\eta$.}
\label{Subrayleigh}
\end{figure}

For $\omega\rightarrow \infty$, the dispersion relation \eq{RW_disp} exhibits the following asymptotic behaviour
\beq
\label{RW_asym}
\mD(m_R,\omega)=\left[(1+\eta)\sqrt{1-2h_0^2m_R^2}-(1-2h_0^2m_R^2+\eta)^2\right]\frac{\omega^{5}\ell^5}{m_R^5c_s^5}+O(\omega^{4}).
\eeq
The minimum value for the normalized surface waves speed, depending on $\eta$ and $h_0$, is given by the value of $m_R$ for which the coefficient 
of the leading order term of \eq{RW_asym} vanishes, and then it can be evaluated by solving the equation:
\beq
\label{RW_lambda}
\Lambda(\eta, h_0, m_R)=(1+\eta)\sqrt{1-2h_0^2m_R^2}-(1-2h_0^2m_R^2+\eta)^2=0.
\eeq
By means of simple algebra, it can be verified that equation \eq{RW_lambda} is equivalent to 
\beq
\label{RW_Ups}
\Upsilon(\eta, h_0, m_R) = \frac{1 - \eta^2 - 2h_0^2m_R^2 + 2\sqrt{1 - 2h_0^2m_R^2}(1 + \eta - h_0^2m_R^2)}{1 + \sqrt{1 - 2h_0^2m_R^2}}=0.
\eeq
The function $\Upsilon$ introduced in expression \eq{RW_Ups} is the same defined in the Wiener-Hopf factorization of steady-state crack propagation problem in \cite{MishPicc1}, 
where the regime $\Upsilon(\eta, h_0, m)> 0$ is studied and a critical limit value for the crack tip speed is individuated by relation \eq{RW_Ups}. Consequently, the minimum
couple stress surface waves propagation velocity coincides with the critical value for steady-state crack propagation, and the condition $\Upsilon(\eta, h_0, m)> 0$ introduced in  \cite{MishPicc1}
defines the transition between two different ranges of velocities, which further in the text will be called sub-Rayleigh and super-Rayleigh propagation regimes. 
These regimes are reported in the $h_0-m$ plane in Fig.\ \ref{Subrayleigh}A. \\ For the case $\eta=0$ the dispersion curves shown in Fig.\ref{dispOm} are identical to that obtained in \cite{MishPicc1} for the shear waves. 
Consequently, for $\eta=0$ the couple stress surface waves degenerate to shear waves and subsonic and sub-Rayleigh regimes are equivalent. This can be demonstrated by the fact that for $\eta=0$ 
the eigenvalue $\beta$ given by (\ref{RW_eigen}) vanishes, and only the term of the matrix \eq{RW_matrix} depending by $\alpha^2$ is non-zero: in that case 
the factor $A$ is also zero and the solution coincides with the planar shear waves solution. 

In Fig.\ \ref{Subrayleigh}A it can be observed that for small values of the rotational inertia the crack propagation is both subsonic and sub-Rayleigh,
and the limit value for the normalized crack tip speed is $m=1$. As $h_0$ increases, the limit speed for sub-Rayleigh regime becomes smaller than for subsonic regime, 
and the critical velocity $m_{c}(h_0, \eta)$ is determined by solving equation \eq{RW_lambda} or alternatively \eq{RW_Ups}. The limit value $h_0^*$ such that for $h_0>h_0^*$ 
the maximum normalized velocity for sub-Rayleigh regime is given by  $m_{c}(h_0, \eta)<1$ is plotted in Fig.\ \ref{Subrayleigh}B as a function of the microstructural parameter $\eta$.

\section{Full-field solution}
\label{sec3}

\begin{figure}[!htcb]
\centering
\hspace{-4mm}
\includegraphics[width=162mm]{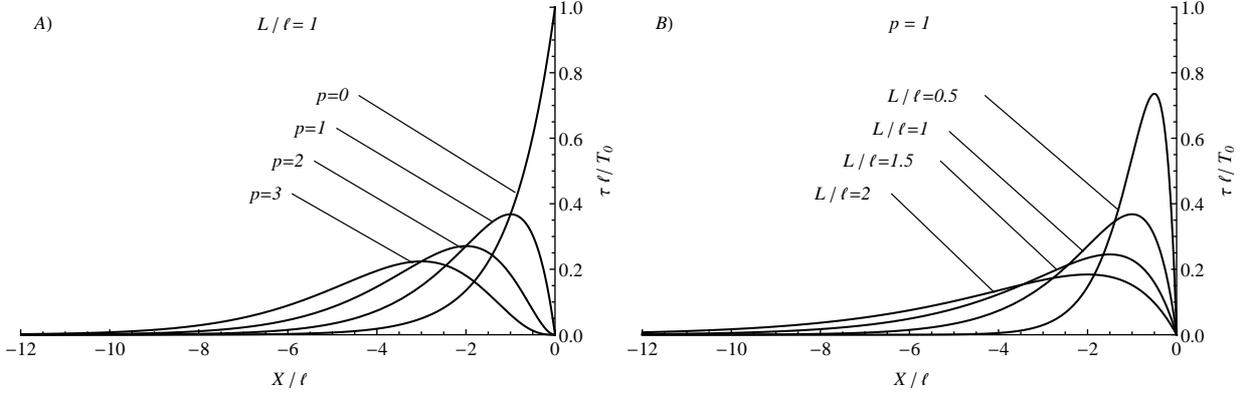}
\caption{\footnotesize Distributed loading applied to the crack faces}
\label{fig}
\end{figure}

The following form for the loading applied on the crack faces is assumed
\beq
\label{load}
\tau(X) = \frac{(-1)^p}{\Gamma(1+p)} \frac{T_0}{L} \left(\frac{X}{L}\right)^p e^{X/L}, \quad X <0, \quad p = 0,1,2,\dots
\eeq
where $\Gamma$ is the Gamma function. It is important to note that the resultant force applied to the upper crack face is $T_0$, indeed
\begin{equation}
\int_{-\infty}^{0} \tau(X) dX = \frac{(-1)^p}{\Gamma(1+p)} \frac{T_0}{L} \int_{-\infty}^{0} \left(\frac{X}{L}\right)^p e^{X/L} dX = T_0.
\end{equation}
Moreover, the maximum of the distributed traction $\tau(X)$ is attained at $X_\text{max} = -pL$. The normalized loading profile $\tau\ell/T_0$ is reported in 
Fig.\ \ref{fig} as a function of $X/\ell$ for several values of the exponent $p$ and of the ratio $L/\ell$. Note that for $p=0$, the loading is bounded but 
different from zero at the crack tip, for $p>0$ the loading tends to zero at the crack tip. Moreover, as $L/\ell$  decreases, the loading is more and more
concentrated around a peak close to the crack tip. 

Sub-Rayleigh regime of propagation defined in previous Section is considered, so that 
\beq
0\ \leq \ m \ \leq \mbox{min}\bigg\{1, m_{c}(h_0, \eta) \bigg\},
\eeq
where the critical value $m_{c}(h_0, \eta)$ is obtained by the solution of equation \eq{RW_lambda} or \eq{RW_Ups} for given values of $\eta$ and $h_0$.

\subsection{Solution of the Wiener-Hopf equation}

Since the Mode III crack problem is skew-symmetric, only the upper half-plane ($y \geq 0$) is considered for deriving the solution. 
The direct and inverse Fourier transforms of the out-of-plane displacements $w(X, y)$ are
\beq
\label{fourier}
\overline{w}(s,y) = \int_{-\infty}^{\infty} w(X,y) e^{i s X} dX, \quad
w(X,y) = \frac{1}{2\pi} \int_{\mL} \overline{w}(s,y) e^{-i s X} ds,
\eeq
respectively, where $s$ is a real variable and the line of integration $\mL$ will be defined later. Applying the Fourier transform (\ref{fourier})$_{(1)}$ to 
equation (\ref{bc1})$_{(1)}$ and using the general factorization procedure illustrated in details in \cite{MishPicc1}, 
the following functional equation of the Wiener--Hopf type can be obtained
\beq
\label{wh}
\overline{p}_3^+(s) + \frac{G\sqrt{s^2\ell^2}}{2\ell}\Psi(s\ell)k(s\ell)\overline{w}^-(s) = \overline{\tau}^-(s),
\eeq
where $\overline{\tau}^-(s)$ is analytic in the lower half complex $s$-plane, $\Im s < 0$ and it is given by
\beq
\label{tauf}
\overline{\tau}^-(s) = \frac{T_0}{(1 + isL)^{1 + p}},
\eeq
where
\beq
k(s\ell) = \frac{1}{\sqrt{s\ell}\Psi(s\ell)(\alpha + \beta)}
\Big\{ \alpha\beta(\alpha^2 + \beta^2 + 2\eta s^2\ell^2) + \alpha^2\beta^2 - \eta^2 s^4\ell^4 \Big\},
\eeq
\beq
\alpha(s\ell) = \sqrt{1 + (1 - h_0^2m^2)s^2\ell^2 + \chi(s\ell)}, \quad
\beta(s\ell) = \sqrt{1 + (1 - h_0^2m^2)s^2\ell^2 - \chi(s\ell)},
\eeq
\beq
\chi(s\ell) = \sqrt{1 + 2(1 - h_0^2)m^2 s^2\ell^2+ h_0^4m^4s^4\ell^4},
\eeq
\beq
\label{Psi}
\Psi(s\ell) = \Upsilon(\eta,h_0,m) s^2\ell^2 + 2\sqrt{1 - m^2},
\eeq
and $\Upsilon(\eta,h_0,m)$ is defined in \eq{RW_Ups}.
The function $k(s\ell)$ has been factorized in \citet{MishPicc1} as $k(s\ell) = k^-(s\ell)/k^+(s\ell)$, where $ s\ell \in \Reals$, and 
$k^+(s\ell)$ and $k^-(s\ell)$ are analytic in the upper and lower half-planes, respectively. Since sub-Rayleigh regime is investigated, $\Upsilon(\eta,h_0, m)$ is positive
for all values of crack tip speed and microstructural parameters considered.

The Wiener-Hopf equation (\ref{wh}) can then be rewritten in the form:
\beq
\label{wh0c2}
\frac{k^+(s\ell)\overline{p}_3^+(s)}{(s\ell)_+^{1/2}} + \frac{G}{2\ell}(s\ell)_-^{1/2} \Psi(s\ell)
k^-(s\ell)\overline{w}^-(s) =
\frac{T_0 k^+(s\ell)}{(s\ell)_+^{1/2}(1 + isL)^{1 + p}},
\eeq
The right-hand side of (\ref{wh0c2}) can be easily split in the sum of plus and minus functions. Indeed,
we use the fact that the function $k^+(s\ell)/(s\ell)_+^{1/2}$ is analytical in the point $sL = +i$ and thus can be represented as
\begin{equation}
\frac{k^+(s\ell)}{(s\ell)_+^{1/2}} = \sum_{j = 0}^{p} (1 + isL)^j F_j + F_{p+1}^+(s)=\sum_{j = 0}^{p} (1 + isL)^j F_j+{\mG}^+(s)(1 + isL)^{p + 1},
\end{equation}
where 
\begin{equation}
{\mG}^+(s) \equiv \frac{F_{p+1}^+(s)}{(1 + isL)^{p+1}} = 
\frac{1}{(1 + isL)^{p + 1}} \left( \frac{k^+(s\ell)}{(s\ell)_+^{1/2}} - \sum_{j = 0}^{p} (1 + isL)^j F_j \right) = O(1),\quad 
s \to +i/L.
\end{equation}
Note that the function ${\mG}^+(s\ell)$ exhibits the following asymptotic behaviour:
\begin{equation}
{\mG}^+(s) =i\frac{F_p}{sL}+O(s^{-2}),\ |s|\to\infty; \quad
{\mG}^+(s) =\frac{k^+(0)}{(s\ell)_+^{1/2}} + O(1), \ |s|\to0,  \quad \mbox{with} \ \Im s>0.\label{ass_0}
\end{equation}
Taking this fact into account, the right-hand side of the equation (\ref{wh0c2})
can be written in the form
\begin{equation}
\label{split}
\frac{T_0 k^+(s\ell)}{(s\ell)_+^{1/2}(1 + isL)^{1 + p}} = T_0{\mG}^-(s) + T_0{\mG}^+(s),
\end{equation}
where
\begin{equation}
{\mG}^-(s) = \sum_{j=0}^{p} \frac{F_j}{(1+isL)^{p+1-j}},
\end{equation}
and
\begin{equation}
\label{Gasym}
{\mG}^-(s) = -i\frac{F_p}{sL}+O(s^{-2}),\ |s| \to \infty; \quad {\mG}^-(s) = \sum_{j = 0}^{p} F_j + O(s),\ |s| \to 0 \quad \mbox{with} \ \Im s<0.
\end{equation}
The unknown constants $F_j$ are computed by evaluating the integrals:
\begin{equation}
\label{fj}
F_j = \frac{L}{2\pi} \oint_{\gamma} \left(\frac{1}{(1 + isL)^{j+1}} \frac{k^+(s\ell)}{(s\ell)_+^{1/2}}\right) ds,
\end{equation}
where $\gamma$ is an arbitrary contour centered at the point $s=i/L$ and lying in the analyticity domain. Substituting (\ref{split}) in (\ref{wh0c2}), we finally obtain:
\begin{equation}
\label{wh0c2_n}
\frac{k^+(s\ell)\overline{p}_3^+(s)}{(s\ell)_+^{1/2}} - T_0{\mG}^+(s) = 
T_0 {\mG}^-(s)-\frac{G}{2\ell}(s\ell)_-^{1/2} \Psi(sl) k^-(s\ell)\overline{w}^-(s).
\end{equation}
The left and right hand sides of (\ref{wh0c2_n}) are analytic functions in the upper and lower half-planes, respectively, and thus define an
entire function on the $s$-plane. The Fourier transform of the reduced force traction ahead of the crack tip and the crack opening gives
$\overline{p}_3^+ \sim s^{1/2}$ and $\overline{w}^- \sim s^{-5/2}$ as $|s| \to \infty$. Therefore, both sides of (\ref{wh0c2_n}) are bounded as 
$|s| \to \infty$ and according to the Liouville's theorem must be equal to a constant $F$ in the entire $s$-plane. As a result, we obtain
\beq
\label{results}
\overline{p}_3^+(s) =
\frac{T_0(s\ell)_+^{1/2}}{k^+(s\ell)} [F + \mG^+(s)],\qquad
\overline{w}^-(s) =
\frac{2T_0\ell}{G}
\frac{\mG^-(s) - F}{(s\ell)_-^{1/2} \Psi(s\ell) k^-(s\ell)}.
\eeq
The constant $F$ is determined by the condition that the displacement $w(X)$ is zero at the crack tip $X = 0$, that is
\beq
\int_{-\infty}^{\infty} \overline{w}^-(s) ds = 0,
\eeq
which leads to
\beq
\label{F}
F =
\frac{\ds \int_{-\infty}^{\infty} \frac{\mG^-(s) ds}{(s\ell)_-^{1/2} \Psi(s\ell) k^-(s\ell)}}
{\ds \int_{-\infty}^{\infty} \frac{ds}{(s\ell)_-^{1/2} \Psi(s\ell) k^-(s\ell)}}={\mG}^-(-i\zeta/\ell),
\eeq
where $\zeta$ is given by
\beq
\zeta=\sqrt{\frac{2\sqrt{1-m^2}}{\Upsilon(\eta,h_0, m)}}.
\eeq
Note here that according to (\ref{ass_0}), $\overline{p}_3^+(0) = T_0$, that is the standard balance condition for this problem. The equivalence between the two alternative 
expressions for the constant $F$ reported in relation \eq{F} can be easily demonstrated by applying the Cauchy integral theorem \citep{Arfk1}.

\subsection{Analytical representation of displacements, stresses and couple stresses}
\label{analex}

The reduced force traction ahead of the crack tip $p_3(X)$ and the crack opening $w(X)$ can be obtained applying the inverse Fourier transform (\ref{fourier})$_2$ to expressions  (\ref{results}).
Since the integrand does not have branch cuts along the real line,
the path of integration $\mL$ coincides with the real $s$-axis. Further, we introduce the change of variable $\xi = s\ell$, thus obtaining
\begin{equation}
w(X) = \frac{T_0}{\pi G} \int_{-\infty}^{\infty}
\frac{\mG^-(\xi/\ell) - F}{\xi_-^{1/2} \psi(\xi) k(\xi) k^+(\xi)}
e^{-iX\xi/\ell} d\xi, \quad X < 0,
\end{equation}
\begin{equation}
p_3(X) = \frac{T_0}{2\pi \ell} \int_{-\infty}^{\infty}
\frac{\xi_+^{1/2} k(\xi)}{k^-(\xi)} [F + \mG^+(\xi/\ell)]
e^{-iX\xi/\ell} d\xi, \quad X > 0.
\end{equation}
The Fourier transform of stress (symmetric and skew-symmetric) and couple stress fields can be derived from (\ref{sigma}), (\ref{tau}) and  (\ref{mu}) namely
\begin{equation}
\overline{\sigma}_{23}(s,0) = -\frac{G}{\ell} \frac{\alpha\beta - \eta s^2\ell^2}{\alpha + \beta} \overline{w}^-(s),
\end{equation}
\begin{equation}
\overline{\tau}_{23}(s,0) = -\frac{G}{2\ell} \frac{1}{\alpha + \beta}
\Big\{\alpha^2\beta^2 + (\alpha^2 + \beta^2 + \alpha\beta)\eta s^2\ell^2 - (1 - 2h_0^2m^2)s^2\ell^2(\eta s^2\ell^2 - \alpha\beta)\Big\} \overline{w}^-(s),
\end{equation}
\begin{equation}
\overline{\mu}_{22}(s,0) = -G (1 + \eta) (is\ell) \frac{\alpha\beta - \eta s^2\ell^2}{\alpha + \beta} \overline{w}^-(s).
\end{equation}
The inverse Fourier transform can be performed as explained above, thus obtaining for $X > 0$
\begin{equation}
\sigma_{23}(X,0) 
= -\frac{T_0}{\pi\ell} \int_{-\infty}^{\infty}
\frac{\alpha(\xi)\beta(\xi) - \eta \xi^2}{\alpha(\xi) + \beta(\xi)}
\frac{\mG^-(\xi/\ell) - F}{\xi_-^{1/2} \psi(\xi) k^-(\xi)}
e^{-iX\xi/\ell} d\xi,
\end{equation}
\begin{align}
\tau_{23}(X,0) 
&= -\frac{T_0}{2 \pi \ell} \int_{-\infty}^{\infty}
\frac{1}{\alpha(\xi) + \beta(\xi)}
\Big\{\alpha^2(\xi)\beta^2(\xi) + (\alpha^2(\xi) + \beta^2(\xi) + \alpha(\xi)\beta(\xi))\eta \xi^2 - \\[3mm]
& \hspace{20mm} {} - (1 - 2h_0^2m^2)\xi^2(\eta \xi^2 - \alpha(\xi)\beta(\xi))\Big\} 
\frac{\mG^-(\xi/\ell) - F}{\xi_-^{1/2} \psi(\xi) k^-(\xi)}
e^{-iX\xi/\ell} d\xi,
\end{align}
\begin{equation}
\mu_{22}(X,0) 
= -\frac{iT_0(1 + \eta)}{\pi}  \int_{-\infty}^{\infty}
\xi \frac{\alpha(\xi)\beta(\xi) - \eta \xi^2}{\alpha(\xi) + \beta(\xi)}
\frac{\mG^-(\xi/\ell) - F}{\xi_-^{1/2} \psi(\xi) k^-(\xi)}
e^{-iX\xi/\ell} d\xi.
\end{equation}

\section{Dynamic energy release rate}
\label{sec4}

In this Section the dynamic energy release rate for a Mode III steady-state propagating crack in couple stress elastic materials under distributed loading conditions given by expression 
(\ref{load}) is evaluated.

\subsection{Explicit evaluation}

The general expression for the dynamic J-integral in couple stress elasticity, including also the rotational inertia contribution, has been derived and proved to be path-independent
in the steady-state case assuming traction free crack faces by \cite{MorPicc1}. Considering the moving framework $OXy$ with the origin 
at the crack tip introduced in Section \ref{sec2}, the J-integral for a steady state crack propagating along the $X-$axis is given by:
\begin{eqnarray}
\mJ & = & \int_{\Gamma}\left[(W+T)n_X-\bp \cdot \frac{\partial\bu}{\partial X}-\bq\cdot \frac{\partial \BGvf}{\partial X}\right]ds=\nonumber\\
    & = & \int_{\Gamma}\left\{(W+T)dy-\left[\bp \cdot \frac{\partial\bu}{\partial X}+\bq\cdot \frac{\partial \BGvf}{\partial X}\right]ds\right\}, \label{Jqp}
\end{eqnarray}
where $\Gamma$ is an arbitrary closed path surrounding the crack tip, and $n_X$ is the Cartesian component directed along the $X-$axis of the outward unit vector normal to $\Gamma$, 
defined by $\bn=(n_X,n_Y,0)$. Since the distributed loading of profile \eq{load} acting on the crack line is assumed, in our case
the contribution of the crack faces must be taken into account, and then in principle the J-integral \eq{Jqp} is not path-independent. 
Nevertheless, in this Section the J-integral is used to determine the dynamic energy release rate evaluating the limit for $\Gamma\rightarrow 0$ in 
\eq{Jqp} \citep{Freund1}. This means that the asymptotic expressions of displacement and stresses can be used for calculating the energy release rate.  
Remembering the asymptotics behaviour of displacement and stresses for antiplane cracks reported in \cite{MorPicc1} and the loading function \eq{load}, 
it is easy to verify that in the limit $\Gamma\rightarrow 0$ the contribution of the crack faces to the J-integral \eq{Jqp} vanishes.

 We assume the rectangular-shaped integration contour $\Gamma$ considered in \cite{MorPicc1}, and in order
to evaluate the energy release rate we allow the height of the path along the $y-$direction to vanish and we make the limit $\Gve\rightarrow 0$. Assuming this type of contour,  
first introduced by \cite{Freund2}, solely asymptotic expressions of displacements and stress fields are required for evaluating the energy release rate. Moreover, upon this choice of path,
allowing the height of the rectangle along the $y-$direction to vanish, the integral $\int_{\Gamma}(W+T)dy$ becomes zero and then the energy release rate is given by 
\beq
\mE=\lim_{\Gamma\rightarrow 0}\mJ=-2\lim_{\Gve\rightarrow 0}\int_{-\Gve}^{\Gve}\left[\bp \cdot \frac{\partial\bu}{\partial X}+\bq\cdot \frac{\partial \BGvf}{\partial X}\right]ds.
\eeq
Since boundary conditions (\ref{bc1}) together with anti-symmetry conditions (\ref{bc2}) provide that the reduced traction $q_1=\mu_{21}$ is zero along the whole crack
$y=0$, the dynamic energy release rate for a steady-state Mode III crack becomes:
\begin{align}
\mE &= -2\lim_{\Gve\rightarrow 0^+}\int_{-\Gve}^{+\Gve}\left\{\left[t_{23}(X,0^+)+\frac{1}{2}\Gm_{22}(X,0^+)\right]\frac{\partial w(X,0^+)}{\partial X}+\Gm_{21}(X,0^+)\frac{\partial \Gvf_1 (X,0^+)}{\partial X}\right\}dX \nonumber \\
    &= -2\lim_{\Gve\rightarrow 0^+}\int_{-\Gve}^{+\Gve}\left[t_{23}(X,0^+)+\frac{1}{2}\Gm_{22}(X,0^+)\right]\frac{\partial w(X,0^+)}{\partial X}dX.\label{J_rect}
\end{align} 
In the limit $|s|\rightarrow \infty$, the Fourier transform of displacements, total shear stress and couple stress fields derived in Section \ref{sec3} assume the following behaviour:
\begin{align}
\overline{w}^-(s,0^+)        &= -\frac{2FT_{0}\ell}{G\Upsilon(h_0,m,\eta)}(s\ell)_-^{-5/2}+O\left((s\ell)_-^{-7/2}\right), \quad 
\Im s<0. \label{four22} \\
\overline{t}_{23}^+(s,0^+)   &= -\frac{FT_{0}(1+\eta-2h_0^2m^2)}{\Upsilon(h_0,m,\eta)}(s\ell)_+^{1/2}+O\left((s\ell)_+^{-1/2}\right), \quad 
\Im s>0,\label{four11} \\
\overline{\mu}_{22}^+(s,0^+) &= \frac{2iFT_{0}\ell\left(\sqrt{1-2h_0^2m^2}-\eta\right)(1+\eta)}{\Upsilon(h_0,m,\eta)\left(1+\sqrt{1-2h_0^2m^2}\right)}(s\ell)_+^{-1/2}+O\left((s\ell)_+^{-1}\right), \quad \Im s>0, \label{four12}
\end{align}
further, we consider the following transformation formula \citep{Roos1}:
\beq
x^{\kappa}\stackrel{ft}{\leftrightarrow}i^{\kappa+1}\Gamma(\kappa+1)s^{-\kappa-1},\ \mbox{with}\ \kappa\neq-1,-2,-3\ldots,
\label{abel}
\eeq
where $\Gamma$ is the gamma function and the symbol $\stackrel{ft}{\leftrightarrow}$ indicates that the quantities on the two sides of the (\ref{abel})  are connected by means of
unilateral Fourier transform. Applying the formula (\ref{abel}) to expressions (\ref{four22})-(\ref{four12}), we get:
\begin{align}
w(X,0^+)          &=  -\frac{8FT_{0}(i\ell)^{-3/2}}{3\sqrt{\pi}G\Upsilon(h_0,m,\eta)}(-X)^{3/2}, \quad X<0. \label{wreal} \\
t_{23}    (X,0^+) &= -\frac{FT_{0}(1+\eta-2h_0^2m^2)(i\ell)^{1/2}}{2\sqrt{\pi}\Upsilon(h_0,m,\eta)}X^{-3/2}, \quad X>0, \label{t23real} \\
\mu_{22}  (X,0^+) &=  \frac{2FT_{0}\left(\sqrt{1-2h_0^2m^2}-\eta\right)(1+\eta)(i\ell)^{1/2}}
{\sqrt{\pi}\Upsilon(h_0,m,\eta)\left(1+\sqrt{1-2h_0^2m^2}\right)}X^{-1/2}, \quad X>0. \label{mu22real}
\end{align}
Then, by substituting expressions (\ref{wreal}), (\ref{t23real}), and (\ref{mu22real}) into equation (\ref{J_rect}), we obtain:
\begin{align}
\mE &= -\frac{4iF^2T_0^2\left[(1+\eta-2h_0^2m^2)+\left(\sqrt{1-h_0^2m^2}-\eta\right)\left(1+\eta\right)\right]}{\pi G\ell \Upsilon^2(h_0,m,\eta)\left(1+\sqrt{1-h_0^2m^2}\right)}
 \lim_{\Gve\rightarrow 0^+}\int_{-\Gve}^{+\Gve}X_{-}^{1/2}X_{+}^{-3/2}dX \nonumber\\
    &= -\frac{4iF^2T_0^2}{\pi G\ell \Upsilon(h_0,m,\eta)}
 \lim_{\Gve\rightarrow 0^+}\int_{-\Gve}^{+\Gve}X_{-}^{1/2}X_{+}^{-3/2}dX 
\label{J_distr}
\end{align}
where $X_{-}^{1/2}$ and $X_{+}^{-3/2}$ are distributions of the bisection type. For any real $\lambda$ with the exception of $\lambda=1,2,3,\dots$,
this particular type of distribution is defined as follows:
$$
X_{+}^{\lambda} = \left\{ \begin{array}{cc}
|X|^\lambda , & \text{for}~ X  > 0, \\
0 , & \mbox{for} ~ X < 0.
\end{array} \right.,\
X_{-}^{\lambda} = \left\{ \begin{array}{cc}
0 , & \mbox{for}~ X > 0, \\
|X|^\lambda, & \text{for}~ X < 0.
\end{array} \right.
$$
The products of distributions inside the integrals in (\ref{J_distr}) is evaluated through the application of Fisher's theorem \citep{Fisch1}, that leads to the relation:
\beq
(X_{-})^\lambda(X_{+})^{-1-\lambda}=-\frac{\pi\delta(x)}{2\sin(\pi\lambda)},\ \mbox{with} \ \lambda\neq-1,-2,-3\dots,
\label{fisher}
\eeq
where $\delta(x)$ is the Dirac delta distribution. Then, by using the relation (\ref{fisher}) into (\ref{J_distr}) and considering the fundamental property of the Dirac delta distribution
$\int_{-\Gve}^{+\Gve}\delta(x)dx=1$, we finally get:
\beq
\mE=\frac{2iF^2T_0^2}{G\ell \Upsilon(h_0,m,\eta)}.
\label{ERR2}
\eeq
A general explicit expression for the dynamic energy release rate associated to an antiplane steady state crack in couple stress elastic materials where a distributed loading of the form 
(\ref{load}) is applied on the crack faces has been derived. Equation (\ref{ERR2}) can be compared with the energy release rate corresponding to a Mode III steady 
state crack in classical elastic materials under the same loading conditions:
\beq
\mE^{cl}=\frac{T_0^2}{GL}\frac{K_p^2}{\sqrt{1-m^2}}, \quad \mbox{with} \quad K_p=\frac{(-1)^p}{p!}\frac{\sqrt{\pi}}{\Gamma(\frac{1}{2}-p)},
\label{ERRclassic}
\eeq
the ratio between the two expressions (\ref{ERR2}) and (\ref{ERRclassic}) is given by 
\beq
\frac{\mE}{\mE^{cl}}=\frac{2iF^2L}{\ell K_p^2 \Upsilon(h_0,m,\eta)}\sqrt{1-m^2}.
\label{ERRratio}
\eeq

\section{Results and discussion}
\label{sec5}

In order to study the effects of loading variations and microstructures on crack propagation, several numerical computations have been performed assuming
loading configurations of the form (\ref{load}) with different values of the exponent $p$ and the ratio $L/\ell$. Total shear 
stress ahead of the crack tip and crack opening profiles are reported and analyzed in subsection \ref{t23w}. Effects of $p$ and $L/\ell$ variation 
on maximum total shear stress ahead of the crack tip and on dynamic energy release rate are discussed in subsections \ref{t23max} and \ref{ERR}, respectively.
The limit cases when the crack tip speed approaches shear waves and couple stress surface waves velocities and 
when the characteristic length $\ell$ vanishes are investigated.

\subsection{Total shear stress and crack opening}
\label{t23w}

\begin{figure}[!p]
\centering
\includegraphics[width=127mm]{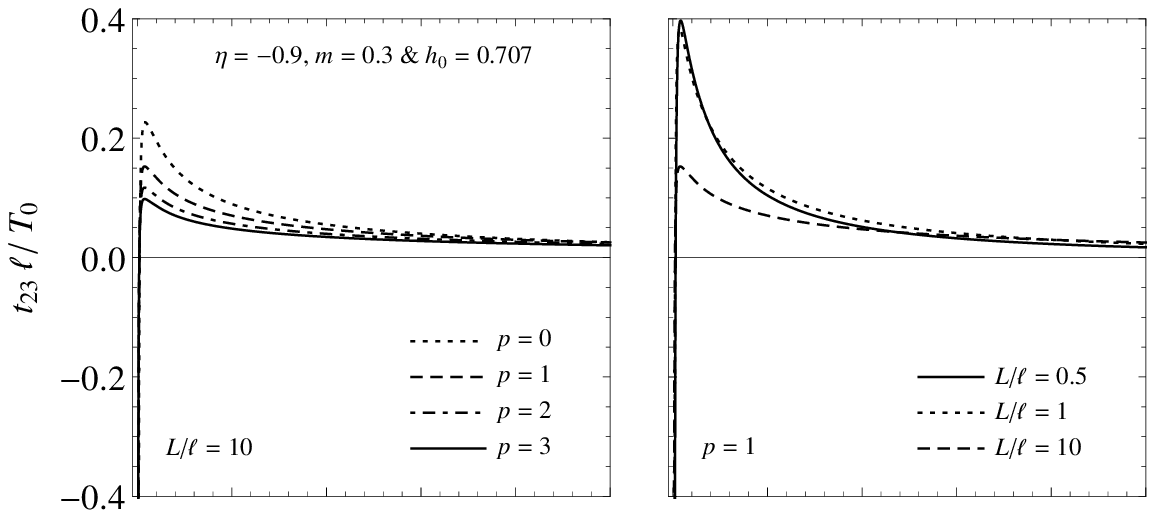}\\[-8mm]
\includegraphics[width=127mm]{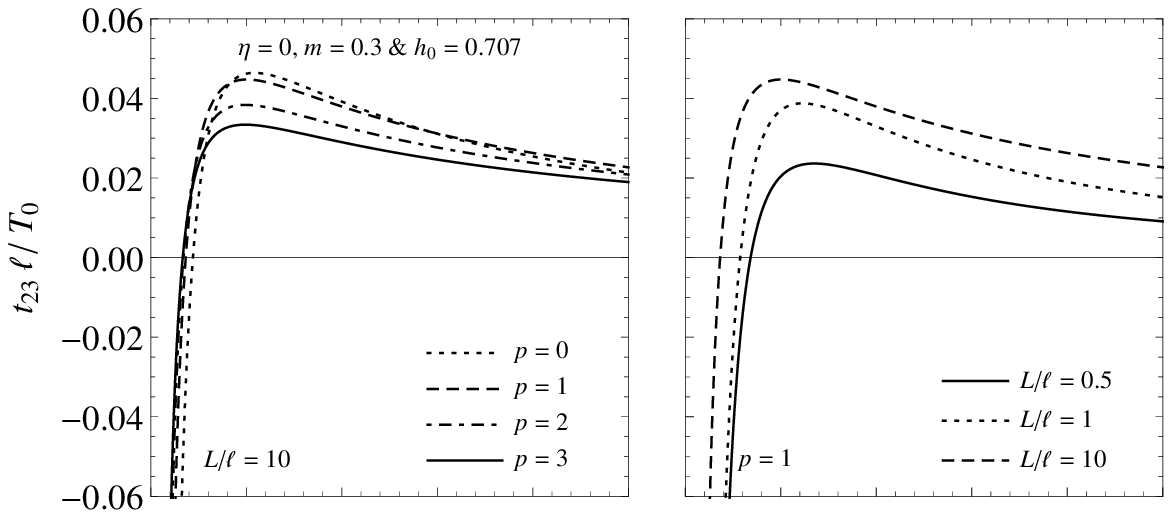}\\[-8mm]
\includegraphics[width=127mm]{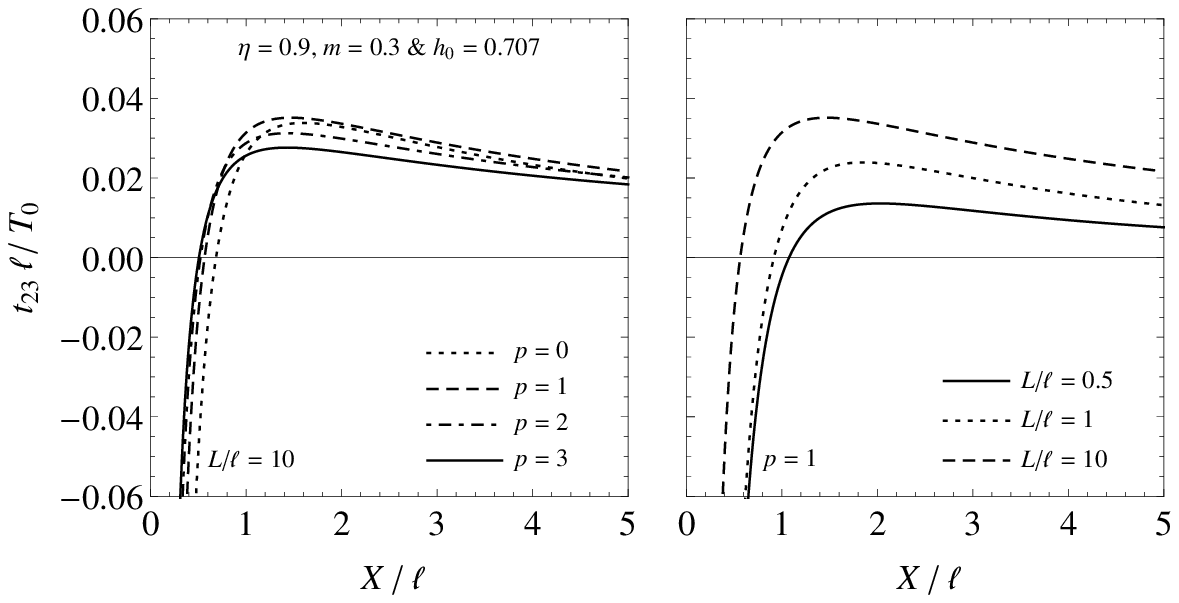}\\[0mm]
\caption{\footnotesize Variation of the total shear stress $t_{23}$ along the $X-$axis.}
\label{figT23}
\end{figure}

In Fig.\ \ref{figT23} the normalized variation of the total shear stress is reported for the same values of the crack tip speed 
$m=0.3$ and of the normalized rotational inertia $h_0=0.707$, and assuming three different values of $\eta=\left\{-0.9, 0, 0.9\right\}$. 

Four different values of $p=\left\{0,\ 1,\ 2,\ 3\right\}$ and three different values of $L/\ell=\left\{0.5, 1, 10\right\}$ have been considered for the computations. 
It can be observed that, as $p$ decreases, and then the maximum of the loading function approaches the crack tip (see Fig.\ \ref{fig}), the level of the shear stress increases. 
This behaviour is more pronounced for $\eta=-0.9$, whereas it becomes less evident for $\eta=0$ and $\eta=0.9$. 
Consequently, for large values of the parameter $\eta$, corresponding to relevant microstructural effects, 
the increasing of the shear stress associate to maximum loading level approaching the crack tip is shielded.

\begin{figure}[!p]
\centering
\includegraphics[width=127mm]{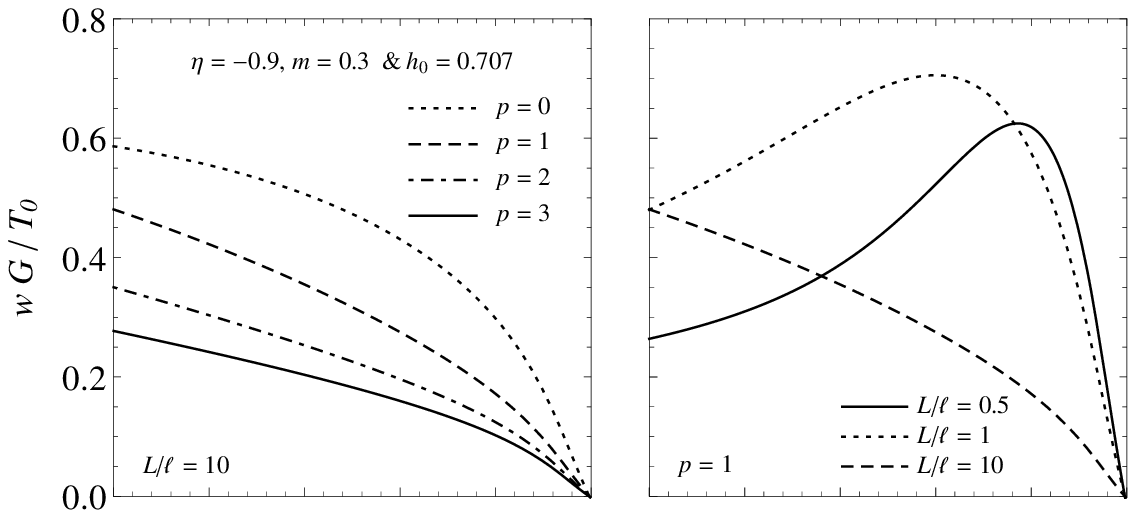}\\[-8mm]
\includegraphics[width=127mm]{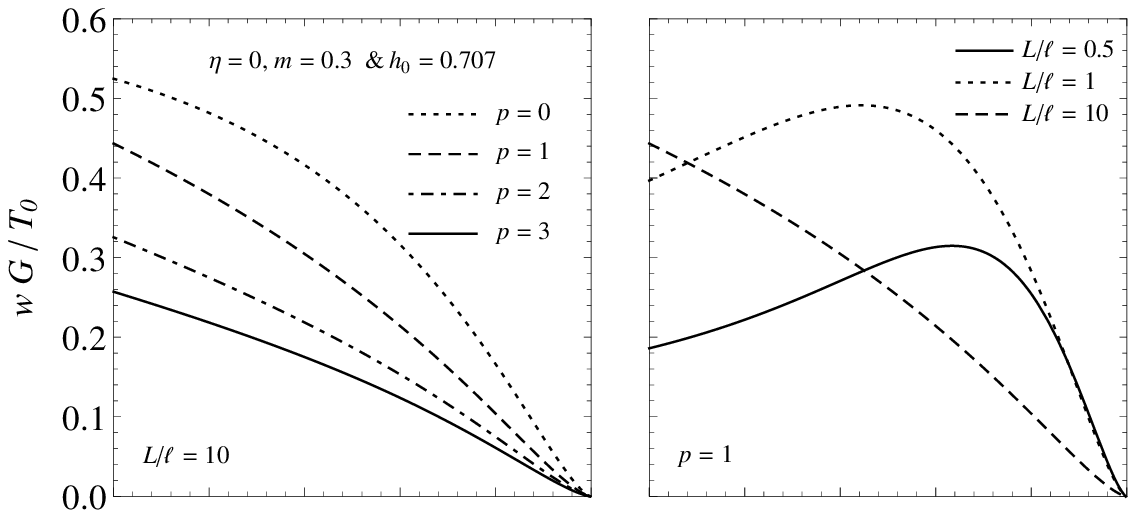}\\[-8mm]
\includegraphics[width=127mm]{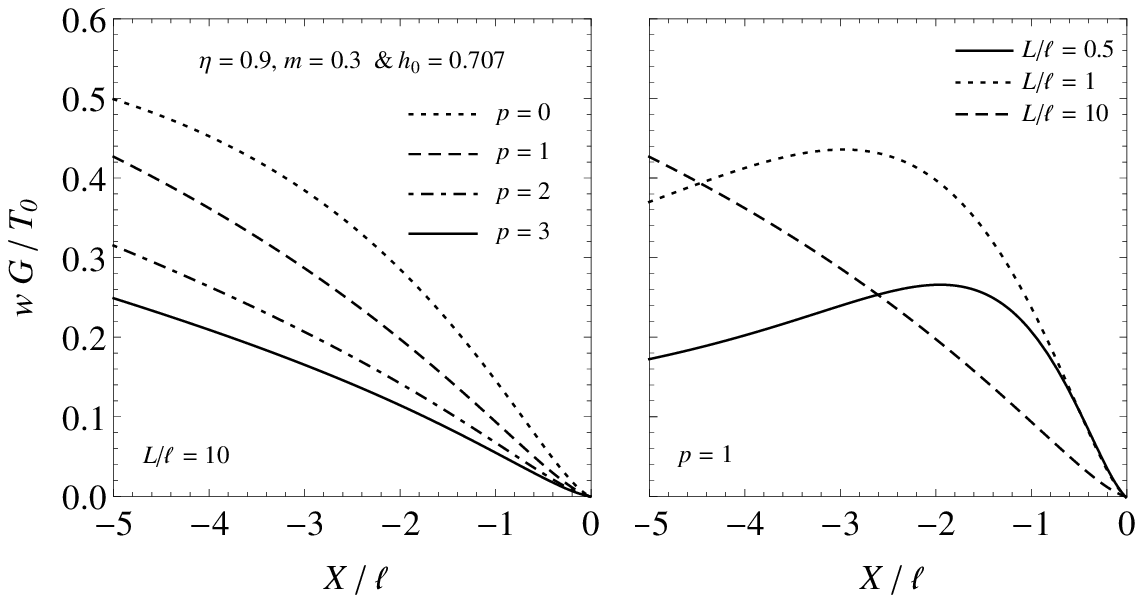}\\[0mm]
\caption{\footnotesize Variation of the crack opening displacement $w$ along the crack faces.}
\label{figW}
\end{figure}

As it is shown in Fig.\ \ref{fig}, small values of the ratio $L/\ell$ correspond to a localization of the applied loading close to the crack tip. In classical elastic 
media, this implies an increasing of the stress level ahead of the crack tip. In presence of couple stress, this increasing is detected for $\eta=-0.9$.
In this case, since $\eta$ is close to the limit value $\eta=-1$, the microstructural effects are not very pronounced and 
the behaviour of the material differs slightly from that of a classical elastic medium \citep{Radi1}. In Fig.\ \ref{figT23}, the increasing of the total shear stress associate 
to the decreasing of the ratio $L/\ell$ is not observed in the cases $\eta=0$ and $\eta=0.9$. 
It means that in couple stress elastic materials, the increasing effect due to the localization of the applied loading is counterbalanced by relevant microstructural contributions, corresponding
to large values of $\eta$. An analogous behaviour is detected for the crack opening in Fig.\ \ref{figW}: the value of $w$ increases as the exponent $p$ decreases and then the maximum of 
the loading function approaches the crack tip, while for small values of $L/\ell$ such as for example $L/\ell=0.5$ the expected increasing of $w$ due to the major localization of the
loading is not observed. Conversely, as the distance from the crack tip increases, the crack opening corresponding to small values of $L/\ell$ approaches a maximum and 
decreases becoming less than in cases where this ratio is greater. This confirms that, as it has been deduced observing total shear stress behaviour ahead of the crack tip, 
the effect of the applied loading localization is shielded by microstructures of the material. \\

\newpage

\subsection{Maximum total shear stress analysis}
\label{t23max}

\begin{figure}[!htcb]
\centering
\includegraphics[width=110mm]{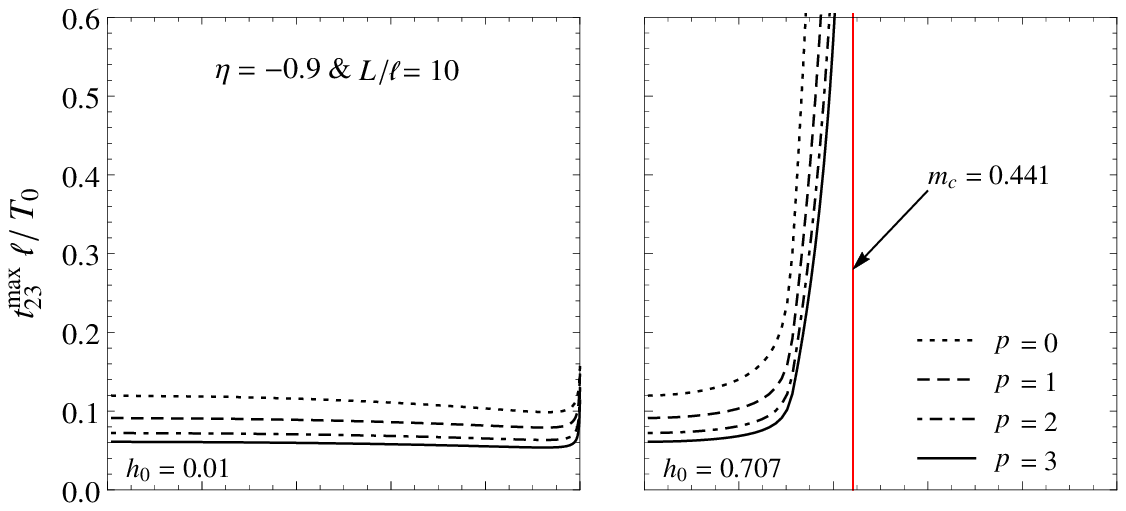}\\[-8mm]
\includegraphics[width=110mm]{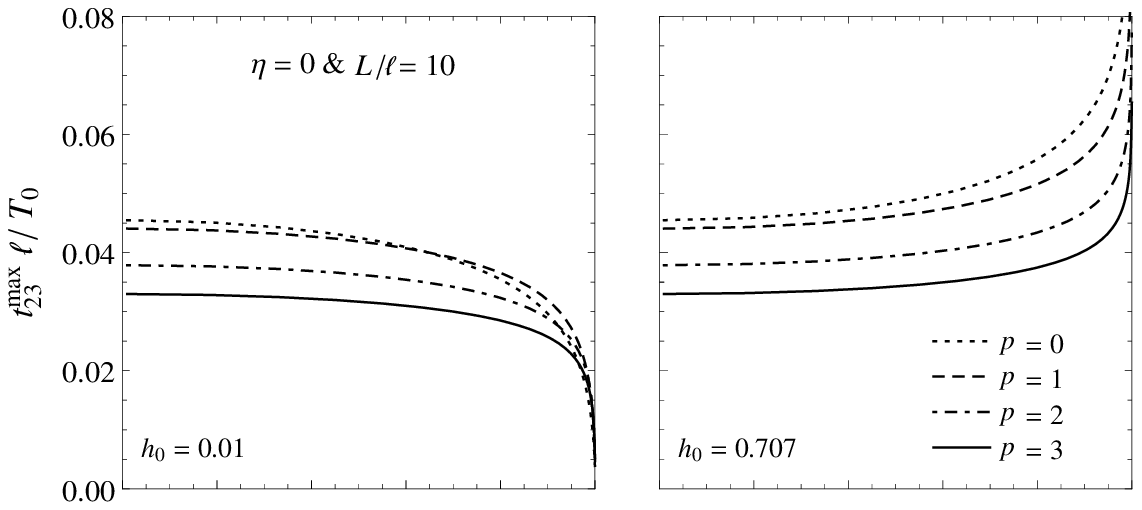}\\[-8mm]
\includegraphics[width=110mm]{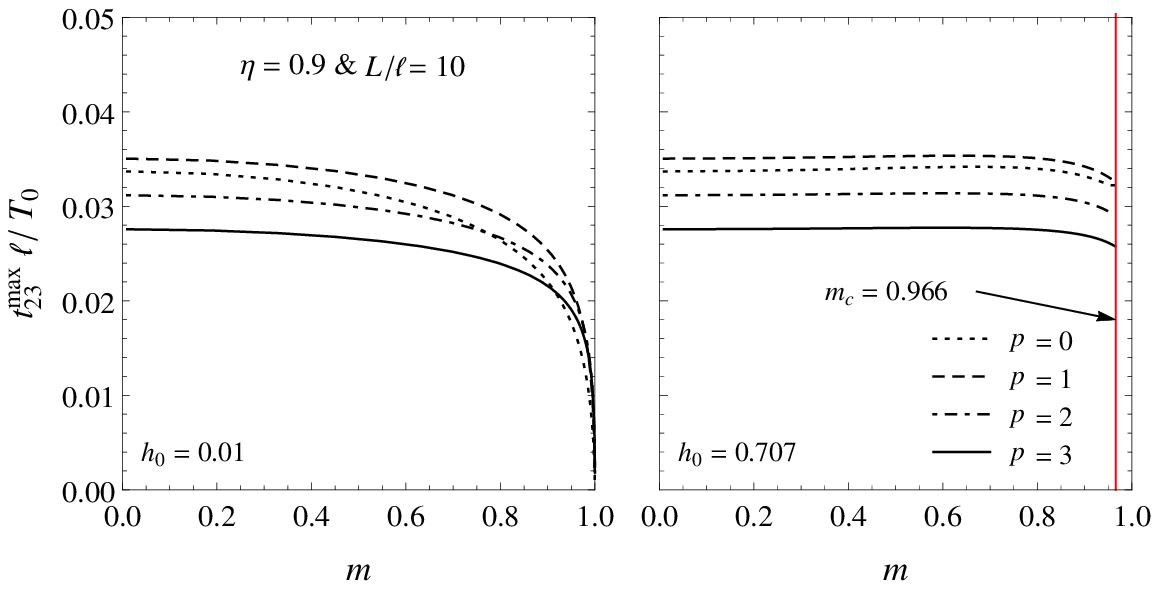}\\[0mm]
\caption{\footnotesize  Variation of the maximum total shear stress $t_{23}^\text{max}$ with the crack tip speed $m$.}
\label{figTmaxm}
\end{figure}

The normalized profile of the maximum total shear stress, $t_{23}^{\text{max}}$, is plotted as a function of the crack tip speed $m$ for
several values of the exponent $p$ and of the rotational inertia $h_0$ in Fig.\ \ref{figTmaxm}. For all sets of parameters considered in the study, numerical results 
show that for the limit cases $m=1$ and $m=m_c$ the maximum total shear stress assumes a finite critical value. 

Observing Fig.\ \ref{figTmaxm}, it can be noted that in the cases $\eta=-0.9$ and $\eta=0$ the level of  $t_{23}^{\text{max}}$ is greater for small values of $p$, 
corresponding to a maximum of the applied loading localized near to the crack tip. Conversely, for $\eta=0.9$ the value of $t_{23}^{\text{max}}$ associated to $p=0$ (dotted lines) is greater 
than for $p=1$ (dashed lines). This is due to the fact that for large values of $\eta$ the presence of the microstructures counterbalances the force action near to the crack tip,
where  the maximum of the loading is applied for $p=0$.

In \cite{Radi1} and \cite{Geo1}, a fracture criterion based on the achieving of a critical level of the maximum shear stress $t_{23}^{\text{max}}=\tau_C$ 
at which the crack starts propagating is defined. Fig.\ \ref{figTmaxm} shows that for $\eta=-0.9$ and $h_0=0.01$ the maximum shear stress decreases as the 
crack speed increases until $m\approx 0.9$, whereas for $m>0.9$ it starts to increase until it reaches the maximum value for $m=1$, when the crack speed approaches the shear waves speed $c_s$.
Differently, for $h_0=0.707$,  $t_{23}^\text{max}$ increases monotonically up to the maximum value corresponding to $m=m_c=0.441$, when the crack tip speed approaches
the minimum velocity for couple stress surface waves propagation. Therefore, referring to the maximum shear stress criterion, for $\eta=-0.9$ and $h_0=0.01$ 
the crack propagation turns out to be initially stable at speed sufficiently lower than the shear wave velocity in classical elastic materials, whereas it becomes unstable when the 
velocity approaches $c_s$. Conversely, for $\eta=-0.9$ and $h_0=0.707$ the propagation is unstable for any $m$ such that $m<m_c$. It can be observed that 
for $\eta=0$ and $h_0=0.01$, $t_{23}^\text{max}$ decreases as the crack tip speed becomes faster and reaches a minimum at $m=1$, while for $h_0=0.707$ it grows as $m$ 
increases until the maximum value corresponding to $m=1$. Consequently, for $\eta=0$ and $h_0=0.01$ the crack propagation can be considered stable, whereas for $\eta=0$ and $h_0=0.707$ 
it turns out to be unstable. On the basis of the same criterion, the figures show that for $\eta=0.9$ the crack propagation 
is stable for both $h_0=0.01$ and $h_0=0.707$. 

The reported results confirm the analysis performed in \citet{MishPicc1}, which shows that relevant microstructural effects, associated to large values of $\eta$, provide a stabilizing effect
of the crack propagation. Moreover, it is important to observe that the variation of the exponent $p$ influences the value of $t_{23}^{\text{max}}$ but not
the qualitative behaviour of its profiles as a function of $m$. This means that if the position of application of the maximum loading is changed, it does not affect the stability of 
the propagation. In Fig.\ \ref{figTmaxm} it can also be noted that for large values of the normalized rotational inertia $h_0$, the level of maximum shear stress
ahead of the crack tip becomes higher. As a consequence, if the contribution of the rotational inertia is not negligible (as for the case $h_0=0.707$), a major amount of energy
must be provided by the loading in order to initiate the propagation and to allow the crack propagating at constant speed.

\begin{figure}[!htcb]
\centering
\includegraphics[width=159mm]{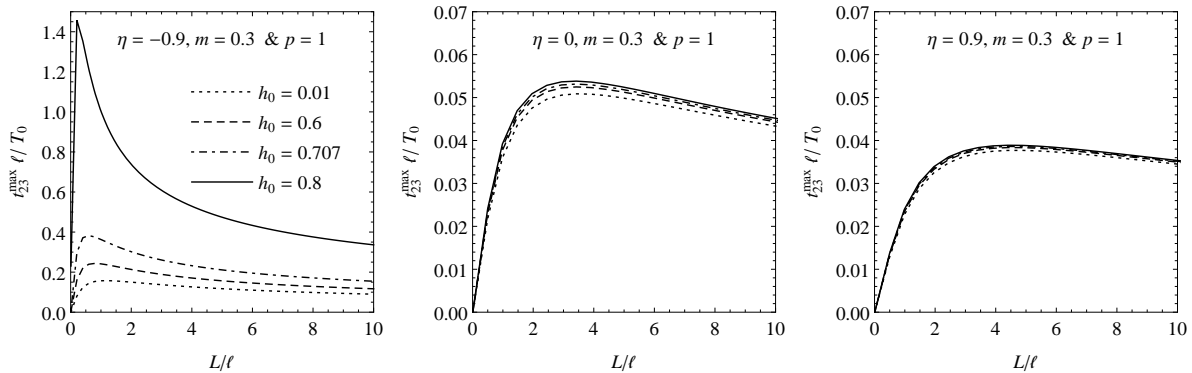}
\caption{\footnotesize Variation of the maximum total shear stress $t_{23}^\text{max}$ with the ratio $L/\ell$.}
\label{figTmaxLtil}
\end{figure}

In Fig.\ \ref{figTmaxLtil} the variation of  $t_{23}^{\text{max}}$ is reported as a function of the ratio $L/\ell$ for $m=0.3, \ p=1$, assuming
$\eta=\left\{-0.9,\ 0, \ 0.9\right\}$ and considering four different values for the normalized rotational inertia $h_0=\left\{0.01,\ 0.6, \ 0.707, \ 0.8\right\}$. 
The decreasing of $L/\ell$, is associated with a strong localization of the applied loading around a maximum close to the 
crack front. As just discussed, in classical elasticity this implies an increasing of the stress level ahead of the crack tip. 
Conversely,  Fig. \ref{figTmaxLtil} shows that in couple stress materials the maximum shear stress is zero for  $L/\ell=0$, then it increases and after reaching a peak it starts decreasing. 
This means that if the loading profile is localized around a maximum close to the crack tip, then its action is shielded by the effects of the microstructure. 
This phenomena is more pronounced for the case $\eta=0.9$, where the microstructural contributions are more relevant. \\
 
\subsection{Energy release rate}
\label{ERR}

\begin{figure}[!htcb]
\centering
\includegraphics[width=160mm]{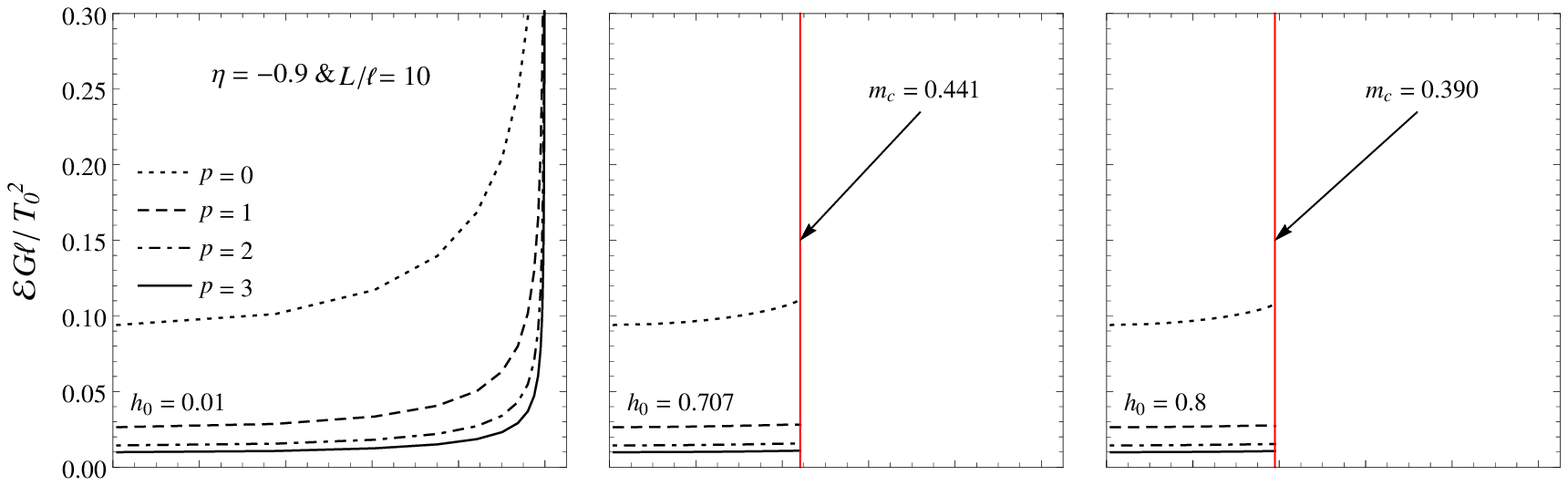}\\[-6mm]
\includegraphics[width=160mm]{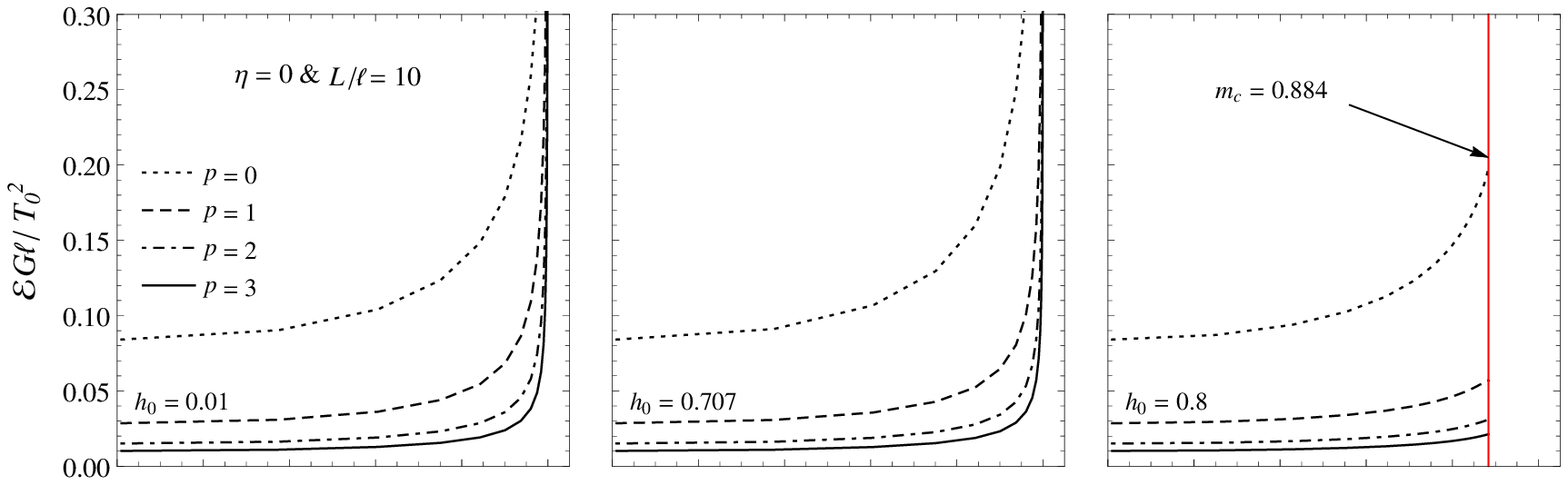}\\[-6mm]
\includegraphics[width=160mm]{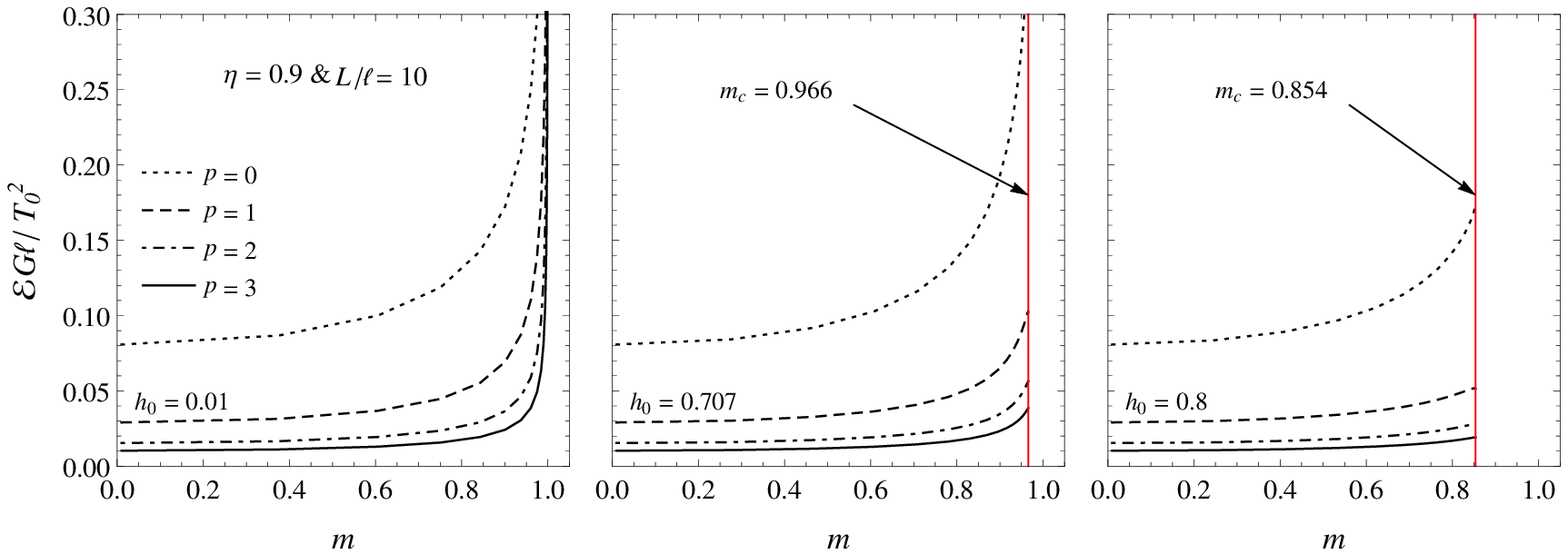}\\[0mm]
\caption{\footnotesize Variation of the energy release rate with the normalized crack tip speed.}
\label{figERRm}
\end{figure}

The normalized variation of the energy release rate versus the crack tip speed $m$ is reported in Fig.\ \ref{figERRm} for the same value of the ratio $L/\ell=10$, 
three different values of $\eta=\left\{0, \ 0.9, \ -0.9 \right\}$ and of the rotational inertia
$h_0=\left\{0.01, \ 0.707,\ 0.8\right\}$. Four different values of $p=\left\{0,\ 1,\ 2,\ 3\right\}$ have been considered in
the computations. The curves reported present the same qualitative behaviour for all values of the exponent $p$:
the energy release rate is initially constant for $m \leq 0.3$, then it increases monotonically until its limiting value corresponding to $m=1$ or $m=m_c$. This means that,
once the critical value $\mathcal{E}_c=2\gamma$ (depending on the material properties) is achieved \citep{Freund1}, the energy release rate always increases as a function of the velocity, 
and then if the applied loading provides the energy necessary for starting the fracture process, the crack has enough energy to accelerate rapidly up to the limiting 
values of the speed \citep{Willis1, ObrMov1}. It follows that, analysing these results by means of Griffith criterion as it has been done in \citet{MorPicc1}, 
the crack propagation turns out to be unstable for any value of the exponent $p$, of the rotational inertia $h_0$ and of $\eta$. Moreover, the variation 
of the loading profile (\ref{load}) does not affect significantly the stability of the propagation, and the stabilizing effect observed for large values of $\eta$ 
applying the $t_{23}^\text{max}$ is not detected. As it has been explained and discussed in details by \citet{MorPicc1}, this discrepancy is due to the fact that the energy release rate 
is evaluated using the term of order $r^{3/2}$ of the asymptotic displacement field, corresponding to the singular shear stress term of order $r^{-3/2}$ (see expressions (\ref{wreal}) 
and (\ref{t23real}) in Section \ref{sec4}). This singular contribution dominates very near to the crack tip, but it is not sufficient to describe accurately the physical behaviour of 
the stresses at few characteristic lengths from the crack tip, where higher order terms of the expansions become important \citep{HancDu1,SmithAya1}.

In Fig.\ \ref{figERRm} it can be observed that in the cases where a rotational inertia greater than the reference value $h_0^*$ defined in Section
\ref{sec2} is considered, the limit value for the energy release rate associated to $m=m_c$ is finite for any set of microstructural parameters. Numerical results show that, also in 
the cases where a small rotational inertia $h_0<h_0^*$ is assumed, the limit maximum value $\mathcal{E}_{\text{max}}$ corresponding to $m=1$ is finite. The only exception is represented
by the case $\eta=0$ and $h_0=1/\sqrt{2}\approx 0.707$: for these particular values of microstructural parameter $\eta$ 
and rotational inertia $h_0$, couple stress surface waves degenerate to non-dispersive shear waves (see dispersion curves in Fig.\ \ref{dispOm}), 
and for $m=1$ the energy release rate becomes unbounded.

\begin{figure}[!htcb]
\centering
\includegraphics[width=160mm]{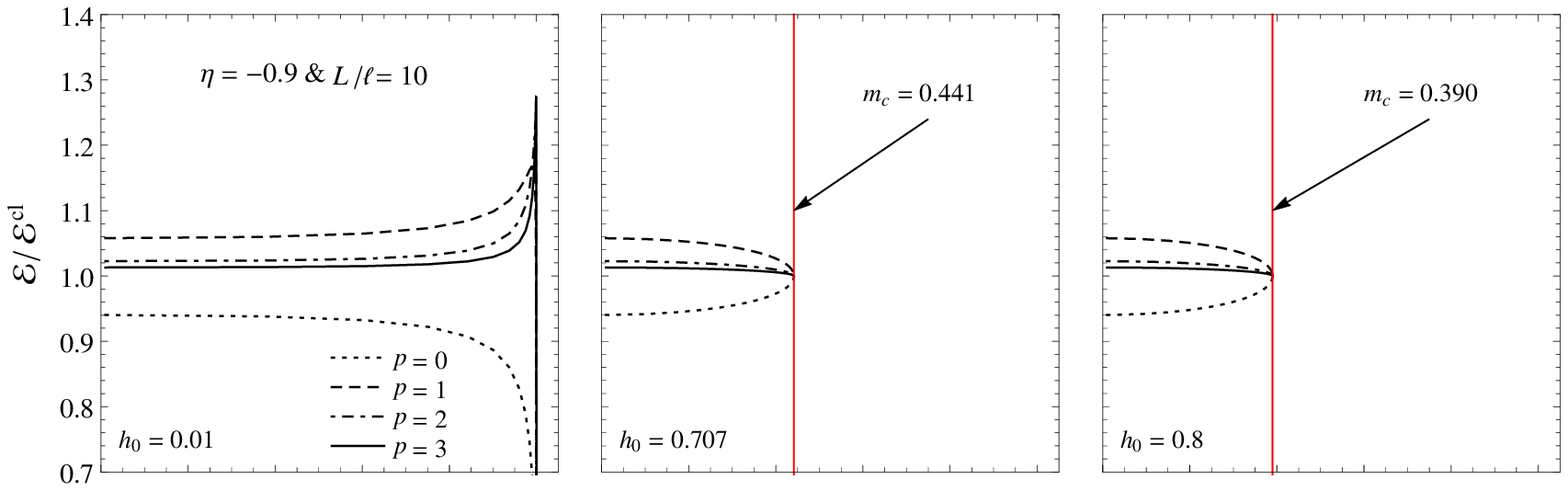}\\[-6mm]
\includegraphics[width=160mm]{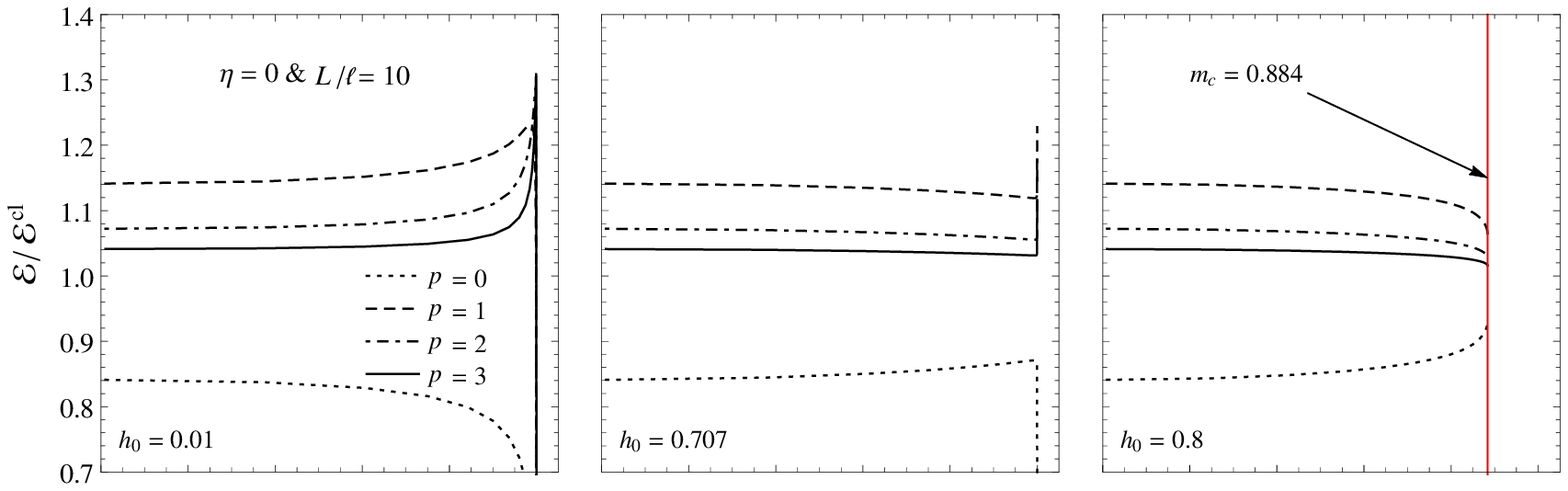}\\[-6mm]
\includegraphics[width=160mm]{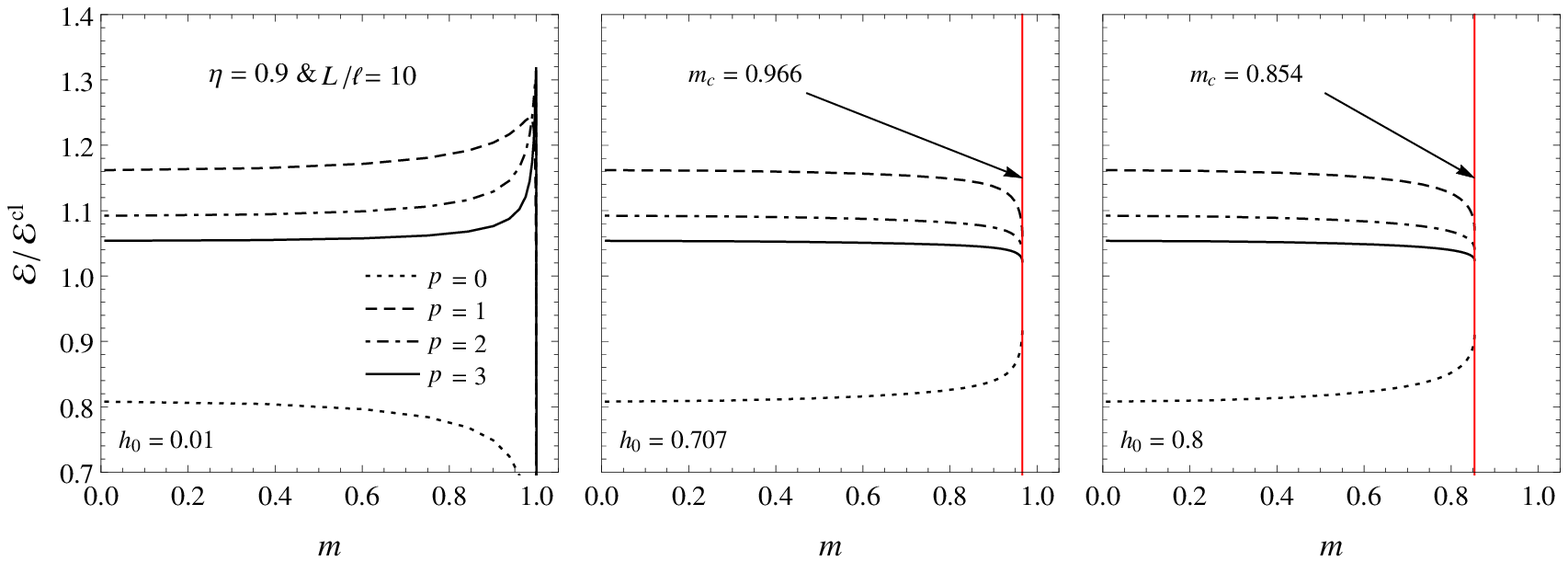}\\[0mm]
\caption{\footnotesize Variation of the ratio $\mathcal{E}/\mathcal{E}^{cl}$ with the normalized crack tip speed.}
\label{figEEcm}
\end{figure}

The ratio between the energy release rate in couple stress materials and the energy release rate in classical elastic materials (\ref{ERRratio}) is plotted 
in Fig.\ \ref{figEEcm} as a function of the normalized crack tip speed $m$. These figures show that 
$\mathcal{E}/\mathcal{E}^{cl}$ is less than one for $p=0$, while it is greater than one for $p>0$. As a consequence, if the maximum of the loading 
is applied at the crack tip, in couple stress elastic materials a minor quantity of energy is provided for propagating cracks at a constant speed 
respect to classical elastic material. This means that for $p=0$ the action of the applied forces is shielded by the effects of the microstructures. Conversely, 
if the maximum loading is not applied at the crack tip, a major amount of energy is available in order to propagate the fracture at 
a given constant velocity respect to classical elastic media. It follows that for $p>0$ the presence of microstructures facilitate the propagation and a weakening 
effect is detected. The observed shielding effect is in agreement with what has been illustrated analyzing crack opening and maximum shear stress for $p=0$.
Note that also in this case both shielding and weakening phenomena are more pronounced for great values of $\eta$, corresponding to relevant microstructural effects. 

Fig.\ \ref{figEEcm} shows that for $h_0=0.01$, and in general for values of the rotational inertia such that $h_0<h_0^*$, 
the ratio $\mathcal{E}/\mathcal{E}^{cl}$ tends to zero at $m=1$. This is due to the fact that while the energy release rate in couple stress materials
reach a finite limit value for $m=1$, in classical elasticity it diverges (see expression \eq{ERRclassic}). The only case where $\mathcal{E}/\mathcal{E}^{cl}$ may reach a non-zero 
value for $m=1$ corresponds to $\eta=0$ and $h_0=1/\sqrt{2}\approx 0.707$. For this particular 
values of $\eta$ and $h_0$, both $\mathcal{E}$ and $\mathcal{E}^{cl}$ becomes unbounded as $m=1$, and then their ratio can be different from zero. Observing  
Fig.\ \ref{figEEcm}, it can also be noted that in all cases where a rotational inertia greater than $h_0^*$ is considered, the ratio
$\mathcal{E}/\mathcal{E}^{cl}$ assumes a finite non-zero limit value for $m=m_c$. In particular, due to the fact that for small values of $\eta$ the microstructural
effects are negligible and the behaviour of the material is similar to that of a classical elastic body \citep{Radi1}, in the case $\eta=-0.9$ the ratio $\mathcal{E}/\mathcal{E}^{cl}$ tends 
to one for $m=m_c$ independently of the value of the exponent $p$. As $\eta$ increases, and then the action of the microstructures becomes relevant,
the difference between the limit values of the ratio associated to different values of $p$ grows.

\begin{figure}[!htcb]
\centering
\includegraphics[width=125mm]{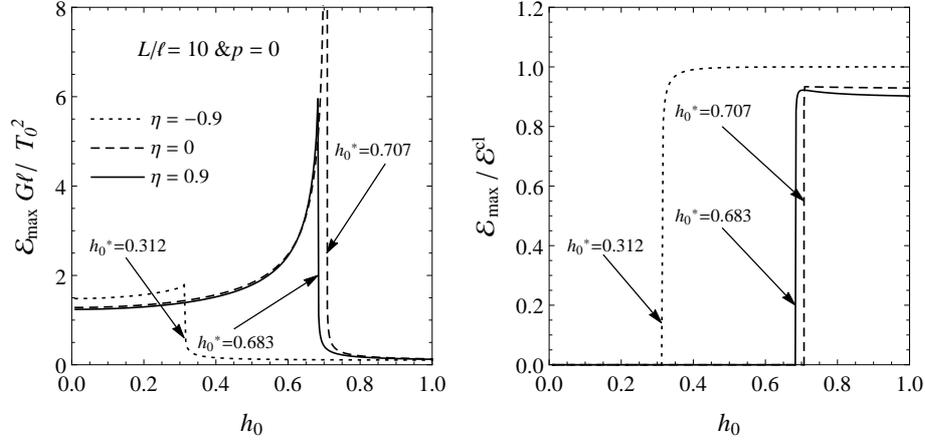}
\caption{\footnotesize Variation of the maximum value of the energy release rate and of the ratio $\mathcal{E}_{\text{max}}/\mathcal{E}^{cl}$ with $h_0$ plotted for $p=0$ and $L/\ell=10$.}
\label{figERRcritic}
\end{figure}

The limit values for the normalized energy release rate and for the ratio  $\mathcal{E}/\mathcal{E}^{cl}$, denominated respectively as  $\mathcal{E}_{\text{max}}$ and
and $\mathcal{E}_{\text{max}}/\mathcal{E}^{cl}$, are reported in Fig.\ \ref{figERRcritic} as functions of $h_0$. As we can expect on the basis of previous considerations,
the limit value for the ratio $\mathcal{E}_{\text{max}}/\mathcal{E}^{cl}$ is zero for $h_0<h_0^*$, and it presents a constant non-zero value for $h_0>h_0^*$. 
In agreement with Fig.\ \ref{figEEcm}, it can be noted that for $\eta=-0.9$ and $h_0>h_0^*$ $\mathcal{E}_{\text{max}}/\mathcal{E}^{cl}\approx 1$.

\begin{figure}[!htcb]
\centering
\includegraphics[width=125mm]{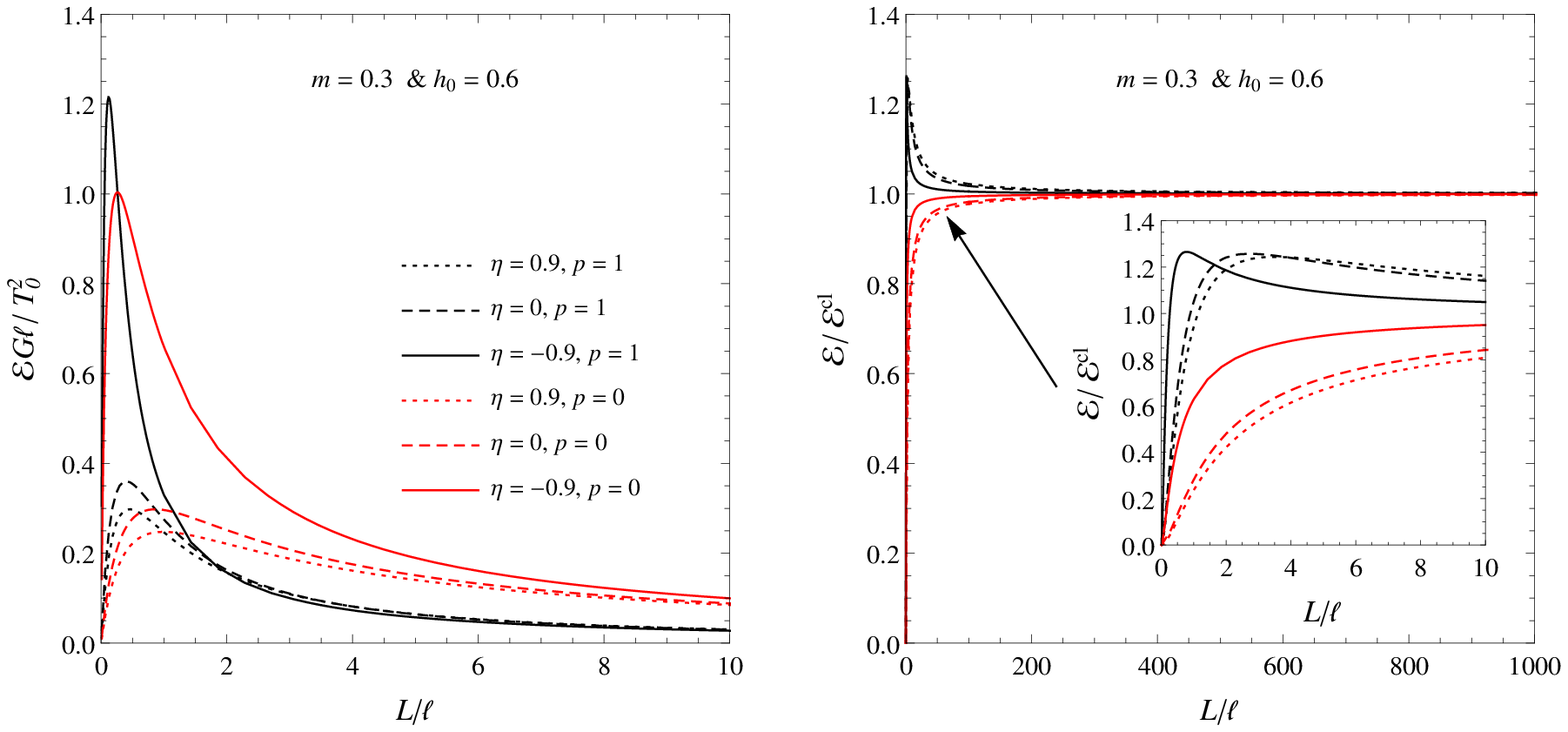}
\caption{\footnotesize Variation of the normalized energy release rate and of the ratio $\mathcal{E}/\mathcal{E}^{cl}$ with $L/\ell$ plotted for $p=1, 2$, $m=0.3$
and $h_0=0.6$.}
\label{figERRLtil}
\end{figure}

In Fig.\ \ref{figERRLtil} the variation of the normalized energy release rate and of the ratio $\mathcal{E}/\mathcal{E}^{cl}$ are plotted as functions of 
$L/\ell$ for $p=1$, $m=0.3$, $h_0=0.707$ and $\eta=\left\{-0.9, \ 0, \ 0.9 \right\}$. The energy release rate tends to zero in the limit 
$L/\ell \rightarrow 0$, then it increases until it reaches a maximum for $L/\ell \approx 0.5$ and then it start decreasing. This behaviour 
means that, due to the shielding effect induced by microstructures, for small values of $L/\ell<0.5$, corresponding to a major localization of 
the applied loading, less energy is provided for propagating the crack at constant speed, and then fracture advancing is hindered.
This shielding effect is also shown by profiles of $\mathcal{E}/\mathcal{E}^{cl}$. Indeed, for $L/\ell<0.5$, if a highly concentrated load is applied close 
to the crack tip in couple stress materials, then $\mathcal{E}/\mathcal{E}^{cl}<1$ and less energy is provided in order to propagate the crack 
with respect to classical elastic media. Differently, for $L/\ell>0.5$ a weakening effect analogous to that observed in Fig.\ \ref{figEEcm}
is detected: $\mathcal{E}/\mathcal{E}^{cl}>1$ and more energy is provided with respect to elastic materials in order to propagate the crack, such that 
crack propagation is favored. It is important to observe that, for all sets of microstructural parameters, as $\ell\rightarrow 0$ and then $L/\ell \rightarrow +\infty$ the 
ratio $\mathcal{E}/\mathcal{E}^{cl}$ tends to one, and the material assumes the classical elastic behaviour. This behaviour is in agreement with the effects observed for plane strain 
problems in \cite{GourGeo2} and \cite{GourGeo1}. It means that as the characteristic scale lengths of the material decrease, couple stress effects becomes negligible, 
and then the material behaviour is identical to that of a classical elastic body for what concerns crack initiation and propagation. This result
is validated by means of the analytical evaluation of the limit of the ratio $\mathcal{E}/\mathcal{E}^{cl}$ as $\ell \rightarrow 0$, reported in the next Section.

\section{Limit of the energy release rate as $\ell \rightarrow 0$ for a general loading function $\tau(X)$}
\label{sec6}

In order to validate the numerical results illustrated in the previous section, the asymptotic behaviour of the dynamic energy release rate \eq{ERR2} as $\ell\rightarrow 0$ is studied. 
For this purpose, the evaluation of the limit of the Liouville constant $F$ as $\ell\rightarrow 0$  is needed. Using explicit expression \eq{F} together with relation \eq{split} 
and Cauchy integral formula, this constant becomes
\beq
\label{Fcauchy}
F={\mG}^-(-i\zeta/\ell)=-\frac{1}{2\pi i T_0}\ds \int_{-\infty}^{\infty} \frac{k^+(s\ell)\overline{\tau}^+(s)}{(s\ell)_+^{1/2} (s+i \zeta/\ell)}ds.
\eeq
Introducing the definition of Fourier transform of the loading function, and remembering that $k^+(z)=1+O(z)$ for $|z|\rightarrow \infty$ (see \cite{MishPicc1} for details), the limit of \eq{Fcauchy} can be written as
\begin{align}
\label{Flim}
\lim_{\ell\rightarrow 0}F &= -\frac{1}{2\pi i T_0}\lim_{\ell\rightarrow 0}\left[\ds \int_{-\infty}^{0} \tau(X)dX \ds \int_{-\infty}^{\infty} 
\frac{e^{isX}}{(s\ell)_+^{1/2} (s+i \zeta/\ell)}ds\right]\nonumber\\
                          &= -\frac{1}{2\pi i T_0}\lim_{\ell\rightarrow 0}\left[\ds \int_{-\infty}^{0} \tau(X)\frac{|X|^{1/2}}{\ell^{1/2}}dX 
\int_{-\infty}^{\infty}\frac{e^{-iy}}{y_+^{1/2} (y+i |X|\zeta/\ell)}dy \right]\nonumber\\
                          &= -\frac{1}{T_0}\lim_{\ell\rightarrow 0}\left[  p \left(i \frac{|X|\zeta}{\ell} \right)\cdot \int_{-\infty}^{0} \tau(X)\frac{|X|^{1/2}}{\ell^{1/2}}dX \right],
\end{align}
where $y=s|X|$. Introducing $t=i|X|\zeta/\ell$, the integral function $p(i|X|\zeta/\ell)$ can be written as 
\beq
\label{pt}
p \left(t \right)= \frac{1}{2\pi i}\int_{-\infty}^{\infty}\frac{e^{-iy}}{y_+^{1/2} (y+t)}dy.
\eeq
For $\ell\rightarrow 0$ and then $|t| \rightarrow \infty$, $p(t)$ exhibits the following asymptotic behaviour
\beq
\label{pt2}
p(t)=\frac{p_1}{t}+O\left(\frac{1}{t^2}\right)=\frac{1}{t\sqrt{\pi}(-i)^{1/2}_{+}}+O\left(\frac{1}{t^2}\right), \quad \mbox{for} \quad |t|\rightarrow +\infty,
\eeq
where $p_1$ is given by the integral 
\beq
\label{p1}
p_1=\frac{1}{2\pi i}\int_{-\infty}^{\infty}\frac{e^{-iy}}{y_+^{1/2}}dy=\frac{1}{\sqrt{\pi}(-i)^{1/2}_{+}},
\eeq
Substituting expression \eq{pt} into the limit \eq{Flim}, it finally becomes
\beq
\label{Flim2}
\lim_{\ell\rightarrow 0}F=-\lim_{\ell\rightarrow 0}\left[\frac{\ell^{1/2}}{\sqrt{\pi}i(-i)^{1/2}_{+}\zeta T_0}\int_{-\infty}^{0} \tau(X)|X|^{-1/2}dX\right]. 
\eeq
Using expression \eq{Flim2}, the limit for $\ell\rightarrow 0$ of the energy release rate \eq{ERR2} can be evaluated:
\begin{align}
\label{ERR_limit}
\lim_{\ell\rightarrow 0}\mE &= \lim_{\ell\rightarrow 0}\frac{2iF^2T_0^2}{G\ell \Upsilon(h_0,m,\eta)} \nonumber \\
                            &= \frac{2}{G\Upsilon(h_0,m,\eta)\pi\zeta^2}\left(\int_{-\infty}^{0} \tau(X)|X|^{-1/2}dX\right)^2 \nonumber \\
                            &= \frac{1}{\pi G\sqrt{1-m^2}}\left(\int_{-\infty}^{0} \tau(X)|X|^{-1/2}dX\right)^2 =\mE_{cl}.
\end{align}
The final result of the limit \eq{ERR_limit} coincides with the definition of energy release rate for a steady-state crack propagating in classical elastic material.
It is important to note that expression \eq{ERR_limit} is valid for any arbitrary loading acting on the crack faces. 
This is in perfect agreement with numerical examples presented in Section \ref{sec5}, which show that $\mE/\mE_{cl}\rightarrow 1 $
for $\ell\rightarrow0$ and then $L/\ell\rightarrow+\infty$. As a consequence, we can say that if $\ell$ and then both characteristic scale lengths $\ell_t$ 
and $\ell_b$ tend to zero, couple stress effects disappear regardless of the applied loading, and then the material behaviour is identical to that of a classical elastic body
for what concerns crack initiation and propagation. 

\section{Conclusions}
The influence of size effects due to microstructures on antiplane dynamic crack propagation in elastic materials is investigated by means 
of indeterminate couple stress theory. Sub-Rayleigh regime for the crack propagation in couple stress media is defined, and the behaviour of the dynamic energy release rate and of the maximum total shear stress is studied considering 
several different loading distributions applied at the crack faces. In the cases where the crack tip speed approaches the shear waves velocity in classical elastic media or
altenatively the minimum couple stress surface waves propagation velocity in the material, a finite limit value for the energy release rate is detected. 
The analysis shows that if the applied loading is localized around a maximum close to the crack tip, its action is shielded by the microstructural effects. 
Conversely, as the profile of the applied loading becomes more uniformly distributed away from the crack tip a greater amount of energy is provided for propagating the crack,
and a weakening effect is observed. Since the predicted shielding and weakening phenomena can strongly influence the level of stress ahead of the crack tip, the analytical results
derived in the present work can represent an important contribution for modelling the mechanical behaviour of microstructured materials.

The asymptotic behaviour of the energy release rate in the limit of vanishing material characteristic lengths is 
studied: numerical examples show that as the microstructural lengths decrease the energy release rate approaches the classical elasticity result. These numerical 
findings are validated by means of a rigorous demonstration. We prove that, independently of the applied loading, in this limit the energy release rate for couple stress materials
tends exactly to the energy release rate for classical elastic materials. This is an important proof of the fact that as the characteristic scale length becomes negligibly small,
size effects vanish and then the material behaviour is identical to that of a classical elastic body for what concerns dynamic crack propagation. 

\section*{Acknowledgements}

L.M. gratefully thank financial support of 
the Italian Ministry of Education, University and Research in the framework of the FIRB project 2010 ``Structural mechanics models for renewable energy 
applications'', 
A.P. and G.M. gratefully acknowledge the support from European Union FP7 projects under contract numbers PCIG13-GA-2013-618375-MeMic and PIAP-GA-2011-286110-INTERCER2, respectively.\\

\newpage

\section*{Appendix A}

\begin{figure}[!htcb]
\centering
\includegraphics[width=10cm]{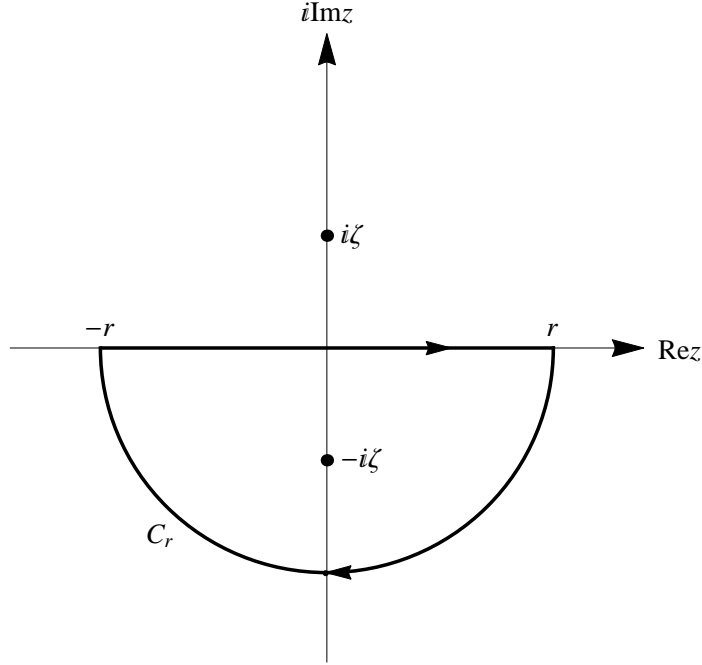}
\caption{\footnotesize Integration path in the complex plane considered for the evaluation of $I_1$ and $I_2$.}
\label{figpath}
\end{figure}

In this Appendix the analytical expression for the Liouville constant \eq{F} is derived. This constant is defined as follows
\beq
\label{Fapp}
F =
\frac{\ds \int_{-\infty}^{\infty} \frac{\mG^-(s) ds}{(s\ell)_-^{1/2} \Psi(s\ell) k^-(s\ell)}}
{\ds \int_{-\infty}^{\infty} \frac{ds}{(s\ell)_-^{1/2} \Psi(s\ell) k^-(s\ell)}}.
\eeq
Commonly, this constant is computed by means of numerical integration procedures. In order to estimate it analytically, we need to calculate explicitly the following two integrals
\begin{align}
I_1 &=  \ds \int_{-\infty}^{\infty} \frac{\mG^-(s) ds}{(s\ell)_-^{1/2} \Psi(s\ell) k^-(s\ell)},\\
I_2 &=  \ds \int_{-\infty}^{\infty} \frac{ds}{(s\ell)_-^{1/2} \Psi(s\ell) k^-(s\ell)}.
\end{align}
These integrals can be represented as limits for $r \rightarrow \infty$:
\begin{align}
I_1 &= \lim_{r \rightarrow\infty} \ds \int_{-r}^{+r} \frac{\mG^-(s) ds}{(s\ell)_-^{1/2} \Psi(s\ell) k^-(s\ell)}= \lim_{r \rightarrow\infty} I_1 (r), \label{I1lim}\\
I_2 &= \lim_{r \rightarrow\infty} \ds \int_{-r}^{+r} \frac{ds}{(s\ell)_-^{1/2} \Psi(s\ell) k^-(s\ell)}= \lim_{r \rightarrow\infty} I_2 (r) \label{I2lim}.
\end{align}
The definite integrals $I_1 (r)$ and  $I_2(r)$ can be evaluated considering the closed integration path in the complex plane illustrated in Fig. \ref{figpath}
\begin{align}
I_1 (r) &= \frac{1}{\ell}\oint_{\Gamma_r} \frac{\mG^-(z/\ell) dz}{z_-^{1/2} \Upsilon(z+i\zeta)(z-i\zeta) k^-(z)}-\frac{1}{\ell} 
  \int_{C_r} \frac{\mG^-(z/\ell) dz}{z_-^{1/2} \Upsilon(z+i\zeta)(z-i\zeta) k^-(z)},\\ 
I_2 (r) &= \frac{1}{\ell}\oint_{\Gamma_r} \frac{dz}{z_-^{1/2} \Upsilon(z+i\zeta)(z-i\zeta) k^-(z)}-\frac{1}{\ell} 
  \int_{C_r} \frac{dz}{z_-^{1/2} \Upsilon(z+i\zeta)(z-i\zeta) k^-(z)},
\end{align}
where $z=s\ell$, and the function $\Psi(z)$ given by expression \eq{Psi} has been decomposed as follows
\beq
\Psi(z) =\Upsilon z^2 + 2\sqrt{1 - m^2}=\Upsilon (z+i\zeta)(z-i\zeta),
\eeq
where $\zeta$ is given by 
\beq
\zeta=\sqrt{\frac{2\sqrt{1-m^2}}{\Upsilon}}.
\eeq
%

Remembering the asymptotic behaviour of the function $\mG^-$ studied in Section \ref{sec3} (see expression \eq{Gasym}) and of $k^-(z)$ reported in \cite{MishPicc1},
it can be easily verified that:
\begin{align}
\lim_{|z| \rightarrow\infty} \frac{z\mG^-(z/\ell)}{z_-^{1/2} \Upsilon(z+i\zeta)(z-i\zeta) k^-(z)} &= 0 \label{cond1}\\
\lim_{|z| \rightarrow\infty} \frac{z}{z_-^{1/2} \Upsilon(z+i\zeta)(z-i\zeta) k^-(z)}              &= 0.\label{cond2} 
\end{align}
Since the conditions \eq{cond1} and \eq{cond2} are satisfied, for the estimation lemma \citep{Arfk1}, the integrals along $C_r$ vanish in the limit $r\rightarrow \infty$
\begin{align}
\lim_{r \rightarrow\infty}\int_{C_r} \frac{\mG^-(z/\ell) dz}{z_-^{1/2} \Upsilon(z+i\zeta)(z-i\zeta) k^-(z)} &= 0,\\ 
\lim_{r \rightarrow\infty}\int_{C_r} \frac{dz}{z_-^{1/2} \Upsilon(z+i\zeta)(z-i\zeta) k^-(z)}               &= 0.
\end{align}
and then the integrals \eq{I1lim} and \eq{I1lim} can be evaluated using Cauchy integral formula \citep{Roos1}. Since the only singularity contained in the 
integration contour is the one at $z=-i\zeta$, the final result is 
\begin{align}
I_1 &= \frac{1}{\ell}\lim_{r \rightarrow\infty}\oint_{\Gamma_r} \frac{\mG^-(z/\ell) dz}{z_-^{1/2} \Upsilon(z+i\zeta)(z-i\zeta) k^-(z)}=\frac{\pi}{\ell\Upsilon}
\frac{\mG^-(-i\zeta/\ell)}{(-i\zeta)_-^{1/2}\zeta k^-(-i\zeta)},\\
I_2 &= \frac{1}{\ell}\lim_{r \rightarrow\infty}\oint_{\Gamma_r} \frac{dz}{z_-^{1/2} \Upsilon(z+i\zeta)(z-i\zeta) k^-(z)}=\frac{\pi}{\ell\Upsilon}
\frac{1}{(-i\zeta)_-^{1/2}\zeta k^-(-i\zeta)}.
\end{align}
The analytical expression for the constant $F$ is finally obtained by the ratio between $I_1$ and $I_2$:
\beq
F=\frac{I_1}{I_2}=\mG^-(-i\zeta/\ell)
\eeq

\section*{Appendix B}

In this Appendix we derive the expression (\ref{ERRclassic}) for the energy release rate corresponding to a Mode III steady state propagating crack in a 
classical isotropic elastic material. For antiplane dynamical problems in classical elasticity the equation of motion (\ref{motion}) becomes
\beq
G\GD u_3= \Gr\ddot{u}_{3}.
\label{eq_motion_cl}
\eeq
Since we are interested in studying steady state crack propagation along $x_1-$axis, we perform the trasformation $u_3(x_1,x_2,t)=w(X,y)$ where
$X=x_1-Vt, y=x_2$, (it is the same substitution illustrated in Section \ref{sec2}), and the  (\ref{eq_motion_cl}) then becomes:
\beq
(1-m^2)\frac{\partial^2 w}{\partial X^2}+\frac{\partial^2 w}{\partial y^2}=0,
\label{eq_motion_clw}
\eeq
where $m=v/c_{s}$ and $c_{s}=\sqrt{G/\Gr}$. The Cauchy stresses are given by
\beq
\Gs_{13}=G\frac{\partial w}{\partial X}, \quad \Gs_{23}=G\frac{\partial w}{\partial y}.
\label{cauchy_cl}
\eeq
The following conditions, equivalent to those imposed for couple stress materials (see equations (\ref{bc1}) and (\ref{bc2})), are assumed 
on the crack surface, at $y=0$:
\begin{align}
\Gs_{23}(y=0) &= -\tau(x), \quad -\infty<x<0, \label{bound_cl-}\\
w(y=0)        &= 0,      \quad  0<x<+\infty, \label{bound_cl+}
\end{align}
where the same distributed loading configuration (\ref{load}) considered for couple stress materials is applied at the crack faces.

An exact solution of the boundary value problem formulated can be obtained by means of Fourier transform and Wiener-Hopf technique. 
The direct and inverse Fourier transform of an arbitrary function $f(x)$ is defined as follows:
\beq
\overline{f}(s,y)=\int_{-\infty}^{+\infty}f(x,y)e^{isx}dx, \quad f(s,y)=\frac{1}{2\pi}\int_{L}\overline{f}(s,y)e^{-isx}ds,
\label{fourier_cl}
\eeq
where $L$ denotes the inversion path within the region of analyticity of the function $\overline{f}(s,y)$ in the complex $s-$plane. 
Transforming the evolution equation (\ref{eq_motion_clw}) we obtain the following ODE:
\beq
\overline{w}''-s^2(1-m^2)\overline{w}=0,
\label{eq_clw_fou}
\eeq
where the prime symbol denotes the total derivative with respect to $y$. The equation (\ref{eq_clw_fou}) possesses the following general 
solution that is required to be bounded as $y\rightarrow+\infty$:
\beq
\overline{w}(s,y)=B(s)e^{-\alpha(s)y},
\label{w_fou}
\eeq
where $\alpha(s)=\sqrt{s^2(1-m^2)}$. The transformed stresses are given by:
\beq
\overline{\Gs}_{13}=-isG\overline{w}, \quad \overline{\Gs}_{23}=G\overline{w}'.
\label{cauchy_cl_fou}
\eeq
The Fourier transforms of the unknown stress ahead of the crack tip $\Gs_{23}(x>0,y=0)$ and of the crack faces displacements $w(x<0,y=0)$ are defined as follows:
\beq
\GS_{23}^+(s)= \int_{0}^{+\infty}\Gs_{23}(x,y=0)e^{isx}dx,
\label{Sigma_fou}
\eeq
\beq
\Gs_{23}(x,y=0) = \frac{1}{2\pi}\int_{D}\GS_{23}^+(s)e^{-isx}ds, \quad x>0,
\label{Sigma_inv}
\eeq
and
\beq
W^-(s)= \int_{-\infty}^{0}w(x,y=0)e^{isx}dx,
\label{W_fou}
\eeq
\beq
w(x,y=0) = \frac{1}{2\pi}\int_{D}W^-(s)e^{-isx}ds, \quad x<0,
\label{W_inv}
\eeq
where the inversion path is assumed to lie inside the region of analyticity of each transformed function. The transformed stress $\GS_{23}^+(s)$ is 
analytic and defined in the lower half complex $s-$plane, $\mbox{Im}s<0$, whereas the transformed displacement $W^-(s)$ is analytic and defined 
in the upper half complex $s-$plane, $\mbox{Im}s>0$.

Taking into account (\ref{w_fou}), and substituting this expression into the (\ref{cauchy_cl_fou})$_{(2)}$, in the limit $y\rightarrow 0$ we obtain:
\beq
B(s)=W^-(s), \quad \GS_{23}^+(s)=-\alpha(s)GW^-(s).
\label{SW}
\eeq
As a consequence, equation (\ref{SW}) together with the condition (\ref{bound_cl-}) provides the following Wiener-Hopf equation connecting the two
unknown functions $\GS_{23}^+(s)$ and  $W^-(s)$:
\beq
\GS_{23}^+(s)-\overline{\tau}^-(s)=-s_+^{1/2}s_-^{1/2}\nu GW^-(s),
\label{WH1}
\eeq
where $\nu=\sqrt{1-m^2}$, $\overline{\tau}^-(s)$ is the Fourier transform of the loading function (\ref{load}), defined by expression (\ref{tauf}),
and the function $\sqrt{s^2}$ is factorized as follows \citep{MishPicc1}:
\beq
\sqrt{s^2}=s_+^{1/2}s_-^{1/2},
\label{s_pm}
\eeq
where the functions $s_+$ and $s_-$ are analytic in the upper and in the lower half plane, respectively. Equation \eq{WH1} can then be rewritten as
\beq
\frac{\GS_{23}^+(s)}{s_+^{1/2}}+s_-^{1/2}\nu GW^-(s)=\frac{T_0}{s_+^{1/2}(1+isL)^{1+p}}.
\label{WH2}
\eeq
The right-hand side of the Wiener-Hopf equation (\ref{WH2}) can be split in the sum of plus and minus functions. Indeed, since the function $s_{+}^{-1/2}$ is 
analytical in the point $s=i/L$, it can be represented as follows
\beq
\frac{1}{s_+^{1/2}}=\sum_{j = 0}^{p} (1 + isL)^j H_j + H_{p+1}^+(s)=\sum_{j = 0}^{p} (1 + isL)^j H_j+{\mI}^+(s)(1 + isL)^{p + 1}
\label{s_dev}
\eeq
where
\beq
{\mI}^+(s) \equiv \frac{H_{p+1}^+(s)}{(1 + isL)^{p+1}} = 
\frac{1}{(1 + isL)^{p + 1}} \left( \frac{1}{s_+^{1/2}} - \sum_{j = 0}^{p} (1 + isL)^j H_j \right).
\label{s_rest}
\eeq
The function ${\mI}^+(s)$ exhibits the following asymptotic behaviour:
\begin{equation}
{\mI}^+(s) =i\frac{H _p}{sL}+O(s^{-2}),\ |s|\to\infty; \quad
{\mI}^+(s) =\frac{1}{s^{1/2}} + O(1), \ |s|\to 0, \quad \mbox{with} \ \Im s>0. \label{ass_rest}
\end{equation}
therefore, the right-hand side of the equation (\ref{WH2}) can be written in the form
\begin{equation}
\label{split_cl}
\frac{T_0}{s_+^{1/2}(1 + isL)^{1 + p}} = T_0{\mI}^-(s) + T_0{\mI}^+(s),
\end{equation}
where
\begin{equation}
\label{I-}
{\mI}^-(s) = \sum_{j=0}^{p} \frac{H_j}{(1+isL)^{p+1-j}},
\end{equation}
and
\begin{equation}
{\mI}^-(s) = -i\frac{H_p}{sL}+O(s^{-2}),\ |s| \to \infty; \quad {\mI}^-(s) = \sum_{j = 0}^{p} H_j + O(s),\ |s| \to 0, \quad \mbox{with} \ \Im s<0. 
\end{equation}
The coefficients $H_j$ can be computed analitically applying the definition of generalized derivative of a function $s^\alpha$ to the case $\alpha=-1/2$:
\beq
H_j=\frac{(-1)^j}{j!}\frac{\sqrt{\pi}}{\Gamma\left(\frac{1}{2}-j\right)}\left(\frac{i}{L}\right)^{-1/2}.
\label{Hj}
\eeq
It has been verified that for any $p$ expression (\ref{Hj}) is equivalent to the following integral definition, analogous to the (\ref{fj}) introduced 
for solving the same crack problem in couple stress materials:
\begin{equation}
\label{Hj2}
H_j = \frac{L}{2\pi} \oint_{\gamma} \left(\frac{1}{(1 + isL)^{j+1}} \frac{1}{s_+^{1/2}}\right) ds,
\end{equation}
where $\gamma$ is an arbitrary contour centered at the point $s=i/L$ and lying in the analyticity domain.
Using decomposition (\ref{split_cl}), the Wiener-Hopf equation (\ref{WH2}) becomes 
\beq
\frac{\GS_{23}^+(s)}{s_+^{1/2}}-T_0\mI^+(s)=-s_-^{1/2}\nu GW^-(s)+T_0\mI^-(s)\equiv E(s).
\label{WH_dec2}
\eeq
The functional equation \eq{WH_dec2} defines the function $E(s)$ only on the real line. In order to evaluate this function, it is first necessary 
to examine the asymptotic behaviour of the functions
$\GS_{23}^+(s)$ and $W^-(s)$. It has been demonstrated that for $X\rightarrow0\pm$ the stress and the displacement along the crack 
faces exhibit the following behaviour:
\begin{align}
\Gs_{23}(X,y=0) &= O(X^{-1/2}) \ \mbox{as} \ X\rightarrow 0+,\label{asym1}\\
w(X,y=0)        &= O(X^{1/2})  \ \mbox{as} \ X\rightarrow 0-.\label{asym2}
\end{align}
Following the same procedure illustrated for couple stress materials, expressions (\ref{asym1}) and (\ref{asym2}) can be transformed
applying Abel-Tauper type theorems \citep{Roos1}:
\begin{align}
\GS_{23}^+(s) &= O(s^{-1/2})  \ \text{as} \ |s|\rightarrow\infty \ \mbox{with} \ \Im s>0  ,\label{asym1_tr}\\
W^-(s)        &= O(s^{-3/2})  \ \text{as} \ |s|\rightarrow\infty \ \mbox{with} \ \Im s<0  .\label{asym2_tr}
\end{align}
Considering the asymptotic behaviour of $\GS_{23}$ and $W^+$ and observing expressions (\ref{s_rest}) and (\ref{I-}), we note that the first member of the
Wiener-Hopf equation (\ref{WH_dec2}) is a bounded analytic function for $\Im s>0$ that is zero as $|s|\rightarrow\infty$, whereas the second member is 
a bounded analytic function for $\Im s<0$ that is also zero as $|s|\rightarrow\infty$. Then, for the theorem of analytic continuation, the two members define one and 
the same analytic function $E(s)$ over the entire complex $s-$plane. Moreover, Liouville's theorem leads to the conclusion that $E(s)=0$. As a consequence, 
the transformed shear stress and displacement are given by:
\begin{align}
\GS_{23}^+(s) &= T_0\mI^+(s)s_+^{1/2}, \ \Im s>0 ,\label{WH_sigma}\\
W^-(s)        &= \frac{T_0\mI^-(s)}{\nu Gs_-^{1/2}}, \  \Im s<0. \label{WH_w}
\end{align}
Evaluating the asymptotic leading term $|s|\rightarrow\infty$ of these expressions, we get:
\begin{align}
\GS_{23}^+(s) &= \frac{iT_0H_p}{L}s_-^{-1/2}+O(s^{-1})  \ \text{as} \ |s|\rightarrow\infty\ \text{with} \ \Im s>0, \label{asym1_sol}\\
W^-(s)        &= -\frac{iT_0H_p}{\nu GL}s_-^{-3/2} + O(s^{-2})\ \text{as} \ |s|\rightarrow\infty \ \text{with} \ \Im s<0, \label{asym2_sol}
\end{align}
applying the transformation formula (\ref{abel}) to the (\ref{asym1_sol}) and (\ref{asym2_sol}) we finally obtain:
\begin{align}
\Gs_{23}(X,y=0) &= \frac{i^{1/2}T_0F_p}{L\sqrt{\pi}}X^{-1/2} = 
\frac{(-1)^p}{p!}\frac{T_0}{\sqrt{L}\Gamma\left(\frac{1}{2}-p\right)}X^{-1/2} \ \mbox{as} \ X\rightarrow 0+,\label{sigma_final}\\
w(X,y=0)        &= -\frac{2i^{-3/2}T_0F_p}{\nu GL\sqrt{\pi}}(-X)^{1/2} = 
\frac{(-1)^p}{p!}\frac{2T_0}{\nu G\sqrt{L}\Gamma\left(\frac{1}{2}-p\right)}(-X)^{1/2} \ \mbox{as} \ X\rightarrow 0-.\label{w_final}
\end{align}
The shear traction expression \eq{sigma_final} can then be used for calculating the stress intensity factor:
\beq
K_{III}^{cl}=\lim_{x \rightarrow 0}\sqrt{2\pi X}\Gs_{23}(X,y=0)=\frac{(-1)^p}{p!\Gamma\left(\frac{1}{2}-p\right)}\sqrt{\frac{2\pi}{L}}T_0.
\eeq
The dynamic J-integral for an antiplane steady state propagating crack is evaluated using the \eq{sigma_final} and \eq{w_final} and performing 
the same procedure illustrated for couple stress materials, choosing a rectangular shaped path surrounding the tip and applying the Fisher theorem:
\beq
\mE^{cl}=\frac{iF_p^2T_0^2}{\nu GL^2}=\frac{T_0^2K_p^2}{GL}\frac{1}{\sqrt{1-m^2}},
\label{J_clapp}
\eeq
where 
\beq
K_p=\frac{(-1)^p}{p!}\frac{\sqrt{\pi}}{\Gamma\left(\frac{1}{2}-p\right)}.
\eeq

\newpage

\bibliography{WF_Bib} 
\bibliographystyle{elsarticle-harv}

\end{document}